\documentclass[a4paper]{cas-dc}

\usepackage[numbers,sort&compress]{natbib}
\usepackage{listings}
\usepackage{bm}
\usepackage{mathtools}
\usepackage{physics}
\usepackage{multirow}
\usepackage{float}

\newfloat{listing}{tbp}{lol}
\floatname{listing}{Listing}

\newcommand{\avg}[1]{\left\langle #1 \right\rangle}
\newcommand{\Pten}{\mathsf{P}}

\newcommand{\calK}{\mathcal{K}}
\newcommand{\Rthree}{\mathbb{R}^3}

\definecolor{codegreen}{rgb}{0,0.6,0}
\definecolor{codegray}{rgb}{0.5,0.5,0.5}
\definecolor{codepurple}{rgb}{0.58,0,0.82}
\definecolor{backcolour}{rgb}{0.95,0.95,0.92}
\lstdefinestyle{mystyle}{
    backgroundcolor=\color{backcolour},
    commentstyle=\color{codegreen},
    keywordstyle=\color{magenta},
    numberstyle=\tiny\color{codegray},
    stringstyle=\color{codepurple},
    basicstyle=\ttfamily\footnotesize,
    breakatwhitespace=false, breaklines=true, captionpos=b,
    keepspaces=true, numbers=left, numbersep=5pt,
    showspaces=false, showstringspaces=false, showtabs=false, tabsize=2
}
\begin{document}

\let\WriteBookmarks\relax

\shorttitle{Performance-portable GPU acceleration of dHybridR}
\shortauthors{B. Ostler et~al.}

\title[mode = title]{Performance-portable GPU acceleration of the hybrid
particle-in-cell code dHybridR}

\author[1]{Bricker Ostler}[orcid=0000-0003-4912-0161]
\cormark[1]
\ead{bostler@uchicago.edu}
\affiliation[1]{
    organization={Department of Physics, The University of Chicago},
    city={Chicago},
    state={IL},
    postcode={60637},
    country={USA}
}

\author[2,3]{Miha Cernetic}[orcid=0000-0002-5088-1745]
\affiliation[2]{
    organization={Department of Astronomy and Astrophysics, The University of Chicago},
    city={Chicago},
    state={IL},
    postcode={60637},
    country={USA}
}
\affiliation[3]{
    organization={Cerebras Systems},
    city={Sunnyvale},
    state={CA},
    postcode={94085},
    country={USA}
}

\author[2,4]{Damiano Caprioli}[orcid=0000-0003-0939-8775]
\affiliation[4]{
    organization={Enrico Fermi Institute, The University of Chicago},
    city={Chicago},
    state={IL},
    postcode={60637},
    country={USA}
}

\cortext[1]{Corresponding author}

\begin{abstract}
Hybrid particle-in-cell simulations are widely used to study kinetic processes in collisionless astrophysical and space plasmas, yet the high computational cost of large-scale three-dimensional runs has largely confined production studies to two dimensions or restricted domains. To address this challenge, we present a performance-portable GPU implementation of the hybrid particle-in-cell code \texttt{dHybridR}. The implementation combines OpenMP target offloading with specialized SYCL kernels for the most computationally expensive operations, while preserving a unified CPU-GPU codebase that supports Intel, AMD, and NVIDIA GPUs. On the exascale supercomputers Aurora and Frontier, \texttt{dHybridR} achieves weak scaling efficiencies of $86\%$ to $97\%$ across 49,152 accelerators, and at 256 particles per cell, its full-node GPU throughput is up to $32\times$ that of the vectorized CPU implementation at approximately $90\%$ less energy per particle-update. To our knowledge, no other hybrid particle-in-cell code has reported GPU performance at this scale, leaving \texttt{dHybridR} uniquely positioned to exploit exascale systems. These advances substantially lower the computational barrier to large-scale three-dimensional hybrid-kinetic simulations of collisionless plasmas.
\end{abstract}

\begin{keywords}
Hybrid particle-in-cell \sep Collisionless plasma \sep GPU acceleration \sep OpenMP offload \sep SYCL \sep Performance portability
\end{keywords}

\maketitle

\section{Introduction}\label{sec:intro}

Many astrophysical plasmas are effectively collisionless: particles evolve according to self-consistent collective electromagnetic fields and wave-particle interactions rather than binary Coulomb collisions. Such physical systems are best modeled by the Vlasov--Maxwell system of equations, describing the evolution of the phase-space distribution function $f(\vb{x},\vb{p},t)$ under the self-consistent evolution of the electromagnetic fields $\vb{E}(\vb{x},t)$ and $\vb{B}(\vb{x},t)$. However, Eulerian Vlasov methods \cite{Palmroth+2018,Juno+2018,Ganse+2023,Valentini+2007,Pezzi+2019} become particularly costly for distributions with extended power-law tails requiring large momentum-space domains. Instead, one often employs the so-called \textit{particle-in-cell} (PIC) method \cite{Hockney1988, Birdsall+1991}, which samples the distribution function using a set of  ``macroparticles'' with the same charge-to-mass ratio as the physical ones. Each of these macroparticles evolves over time according to the equations of motion,

\begin{equation} \label{eq:Lorentz-force-law}
    \dv{\vb{x}_p}{t} = \vb{v}_p \qc \dv{\vb{w}_p}{t} = \frac{q_s}{m_s} \pqty\bigg{\vb{E} + \frac{\vb{v}_p}{c}\cross\vb{B}},
\end{equation}

\noindent where $q_s/m_s$ is the species charge-to-mass ratio, $\vb{E}$ and $\vb{B}$ the electromagnetic fields, $c$ the speed of light, and $\vb{w}_p = \gamma \vb{v}_p$ the proper velocity of a macroparticle with position $\vb{x}_p(t)$, velocity $\vb{v}_p(t)$, and Lorentz factor $\gamma$.\footnote{All physics formulae use Gaussian-cgs units.} Since Eq.~\eqref{eq:Lorentz-force-law} defines the characteristics of the Vlasov equation under self-consistent electromagnetic fields, the time-evolution of these macroparticles solves the Vlasov--Maxwell system, capturing the full kinetic physics at play.

For many problems in high-energy astrophysics, space physics, and laboratory astrophysics, only the ions need to be described kinetically; electrons can instead be approximated as an inertialess, charge-neutralizing fluid, where their dynamics are assumed to obey the conservation of momentum equation \cite{Lipatov2002, Winske+2023}. This is the so-called \textit{hybrid} PIC method that our code \texttt{dHybridR} employs \cite{Gargate+2007, Haggerty+2019}. This approximation enables a significant speedup compared to full PIC: the electric field is obtained via a straightforward algebraic equation, and the resolution requirements are set by the ion inertial length and inverse gyrofrequency rather than their smaller electron counterparts, allowing larger cell widths in the Courant--Friedrichs--Lewy (CFL) condition and subsequently enabling larger timesteps that resolve particle gyration. 

Even so, 3D simulation under the hybrid model remains challenging. Cosmic-ray-driven instability and diffusive shock acceleration simulations require cell lengths that resolve the ion inertial length; box sizes on the order of the cosmic ray gyroradius of the most energetic particles, which grows as acceleration progresses; timesteps that satisfy the CFL constraint for the fastest-moving particles at speed $v_{\max}$, which for cosmic rays is near the speed of light $c$; and resolution of the acceleration process itself, which unfolds over thousands of inverse ion gyrofrequencies \cite{Caprioli+2014}. Moreover, certain physical effects only arise in 3D geometries: the artificial translational symmetry that 2D imposes suppresses cross-field diffusion \cite{Jokipii+1993, Giacalone+1994, Jones+1998}, and 3D hybrid simulations of quasi-perpendicular shocks demonstrate spontaneous acceleration of ions to non-thermal energies, an effect entirely absent from 2D simulations \cite{Orusa+2023, Orusa+2026}. 

Given the computational demands, production simulations have largely remained 2D \cite{Haggerty+2020, Le+2023, Zacharegkas+2024}, with 3D runs forced to make significant compromises like restricting the transverse extent to tens of ion inertial lengths or simulating to durations of order $10^2$ inverse gyrofrequencies. Uncompromised 3D simulations require at least an order of magnitude more compute, which most leadership supercomputers concentrate almost entirely in their GPUs. This motivates a GPU-accelerated \texttt{dHybridR}.

Full PIC codes have been widely ported to GPUs in recent years \cite{Myers+2021, Lee+2025, Bird+2022, Hakobyan+2026}, and several have demonstrated performance across multiple GPU vendors at scale \cite{Fedeli+2022, Hakobyan+2026}. The hybrid-PIC landscape is noticeably less mature on all three counts. As far as we are aware, published efforts to date---\texttt{AMITIS} \citep{Fatemi+2017}, \texttt{Menura} \citep{Behar+2022}, \texttt{Hybrid EPIC-GOD} \citep{Kim+2025}, and an Ohm's-law solver in \texttt{WarpX} \citep{Groenewald+2023}---are few in number, primarily single-vendor, and have reported performance only on NVIDIA hardware up to 64 GPUs.

In this paper, we present a performance-portable GPU implementation of \texttt{dHybridR}, demonstrated on Intel, AMD, and NVIDIA hardware at leadership facility scale. The paper is organized as follows. Section~\ref{sec:code} provides an overview of the hybrid PIC model and the steps that constitute a \texttt{dHybridR} timestep. Section~\ref{sec:cpu} describes CPU-side work that preceded the port, and Sec.~\ref{sec:gpu-implementation} details the GPU implementation. Section~\ref{sec:performance} benchmarks the GPU implementation's scaling, speedup over the CPU code, and energy efficiency, with subsequent application to a 3D parallel collisionless shock. Finally, Sec.~\ref{sec:conclusion} summarizes our results.

\section{Hybrid PIC and the \texttt{dHybridR} timestep}\label{sec:code}

\texttt{dHybridR} \cite{Haggerty+2019} is a massively parallel Fortran hybrid PIC code built to study high-energy astrophysical and space-plasma problems in which a small population of non-thermal ions (cosmic rays) affects the dynamics of an otherwise non-relativistic plasma. It generalizes the Newtonian code \texttt{dHybrid} \cite{Gargate+2007} by the inclusion of relativistic ion dynamics, with many of the simulations cited in Sec.~\ref{sec:intro} run using it \cite{Haggerty+2020, Zacharegkas+2024, Orusa+2023, Orusa+2026}. Before discussing specific kernel implementations and their performance, we briefly describe hybrid PIC and detail the steps required to execute a single \texttt{dHybridR} timestep. 

\subsection{Hybrid PIC} \label{sec:hybrid-pic}

The Vlasov equation describing the evolution of the non-relativistic electron distribution function, $f_e(\vb{x},\vb{v},t)$, is given by

\begin{equation} \label{eq:Vlasov-eq}
    \pdv{f_e}{t} + \vb{v}\vdot \grad_x f_e - \frac{e}{m_e}\pqty{\vb{E} + \frac{\vb{v}}{c}\cross \vb{B}} \vdot \grad_v f_e = 0,
\end{equation}

\noindent where $\vb{E}(\vb{x},t)$ and $\vb{B}(\vb{x},t)$ are the total electromagnetic fields. Defining the $n$\textsuperscript{th} moment of a quantity $X$ by $m_e \int_{\Rthree} \vb{v}^{\otimes n} X \dd[3]{v}$ where $\vb{v}^{\otimes n}$ denotes the $n$\textsuperscript{th} tensor product of $\vb{v}$, the zeroth moment of Eq.~\eqref{eq:Vlasov-eq} yields the conservation of mass equation for electrons,

\begin{equation} \label{eq:continuity}
    \pdv{(m_e n_e)}{t} + \div\pqty{m_e n_e \vb{u}_e} = 0.
\end{equation}

\noindent Combining Eq.~\eqref{eq:continuity} with the first moment of Eq.~\eqref{eq:Vlasov-eq} then yields the conservation of momentum equation for electrons,

\begin{equation} \label{eq:conservation-of-momentum}
    m_e n_e \frac{\mathrm{D} \vb{u}_e}{\mathrm{D}t} = - e n_e \pqty{\vb{E} + \frac{\vb{u}_e}{c}\cross \vb{B}} - \div{\Pten_e}.
\end{equation}

\noindent Here, the left-hand side denotes the inertial term for a parcel of electron fluid using the material derivative $\mathrm{D}/\mathrm{D}t \equiv \partial/\partial t + \vb{u}_e \vdot \grad$, $n_e = \int_{\Rthree} f_e \dd[3]{v}$ is the electron number density, $\vb{u}_e = n_e^{-1} \int_{\Rthree} \vb{v} f_e \dd[3]{v}$ the electron bulk velocity, and

\begin{equation}
\begin{aligned}
    \Pten_e &= m_e \int_{\Rthree} \pqty{\vb{v}-\vb{u}_e}\pqty{\vb{v}-\vb{u}_e} f_e \dd[3]{v} \\
    &= m_e \int_{\Rthree} \vb{v}\vb{v} f_e \dd[3]{v} - m_e n_e \vb{u}_e \vb{u}_e
\end{aligned}
\end{equation}

\noindent the electron pressure tensor.

The hybrid approximation assumes that the electrons are a massless, charge-neutralizing fluid. The charge-neutralizing assumption is a statement of quasi-neutrality, namely 

\begin{equation}
    e n_e = \sum_{s\in\calK} q_s n_s \equiv \rho_{\calK},
\end{equation}

\noindent where $\calK$ denotes the set of all kinetic (non-electron) species. Splitting the current density into fluid and kinetic parts, 

\begin{equation}
    \vb{J} = \sum_{s\in\calK\cup\{e\}} q_s n_s \vb{u}_s = \vb{J}_\calK - e n_e \vb{u}_e,
\end{equation}

\noindent with $\vb{J}_\calK = \sum_{s\in\calK} q_s n_s \vb{u}_s$ additionally yields 

\begin{equation}
    \vb{u}_e = \vb{u}_\calK - \frac{\vb{J}}{\rho_\calK},
\end{equation}

\noindent where $\vb{u}_\calK \equiv \vb{J}_\calK / \rho_\calK$ is a charge-density-weighted bulk velocity for the kinetic species.

In the massless limit $m_e \rightarrow 0$ with $\Pten_e$ held finite, the hybrid model then reduces Eq.~\eqref{eq:conservation-of-momentum} to an ``Ohm's law'' for the electric field,

\begin{equation} \label{eq:Ohms-law}
    \vb{E} = -\frac{\vb{u}_\calK}{c}\cross \vb{B} + \frac{\vb{J}\cross\vb{B}}{c\rho_\calK} - \frac{\div{\Pten_e}}{\rho_\calK}.
\end{equation}

Additionally, we restrict our attention to modes with $\abs{\omega} \ll kc$ so the transverse displacement current in the Amp\`ere--Maxwell equation may be discarded, eliminating electromagnetic radiation from the model while retaining high-frequency modes with low phase speed such as whistler waves \cite{Nielson+1976, Hewett1985}. Since the longitudinal displacement current is zero by quasi-neutrality, we obtain

\begin{equation}
    \vb{J} = \frac{c}{4\pi}\curl{\vb{B}},
\end{equation}

\noindent which may be substituted into Eq.~\eqref{eq:Ohms-law}. One must also assume a closure for $\Pten_e$; often, the pressure tensor is taken to be isotropic, $\Pten_e = P_e \mathsf{I}$ with the scalar $P_e$ evolved in time either via the second moment of Eq.~\eqref{eq:Vlasov-eq} or approximated via an appropriate equation of state. All \texttt{dHybridR} simulations in this paper employ a polytropic equation of state $P_e = Kn_e^{\gamma_e}$ as the closure for polytropic constant $K$ and index $\gamma_e$.

\subsection{A \texttt{dHybridR} timestep}

\begin{figure}[pos=t]
\centering
\includegraphics{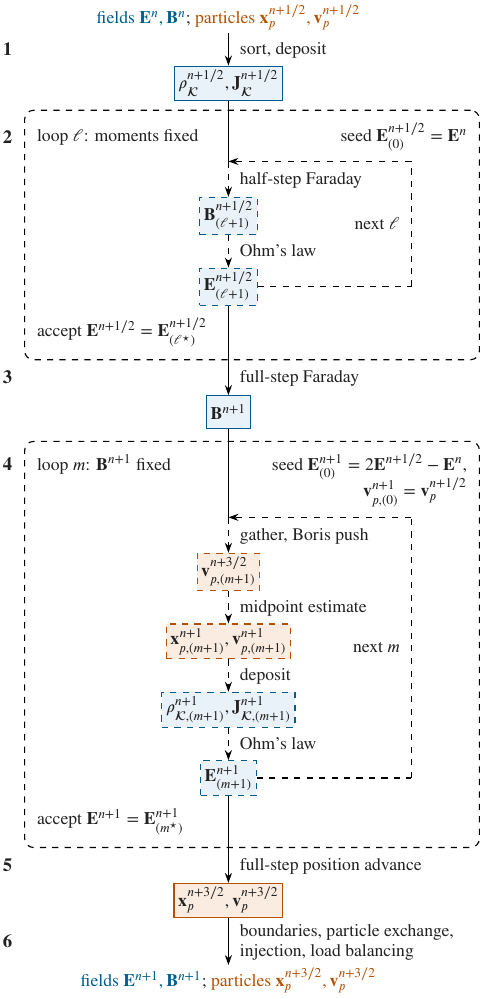}
\caption{\label{fig:timestep-flowchart}Flowchart of one \texttt{dHybridR} timestep, with steps labeled sequentially from 1 to 6. Solid boxes denote accepted values, while dashed boxes denote estimates in a fixed-point loop, whose last iterate is the accepted value. Grid quantities are marked in blue and macroparticle quantities in orange.}
\end{figure}

At time $t^n = n \Delta t$ for integer $n\geq 0$, the electromagnetic fields are given by $\pqty{\vb{E}^{n},\vb{B}^{n}}$ while the particle positions and velocities are given a half-step later, $\pqty{\vb{x}_p^{n+1/2},\vb{v}_p^{n+1/2}}$. A timestep updates these quantities to $t^{n+1}$ and $t^{n+3/2}$, respectively. A flowchart of the timestep is given in Fig.~\ref{fig:timestep-flowchart}. 

As in full PIC, electromagnetic fields reside on a background grid and macroparticles are assigned a certain \textit{shape function} $S$ that determines how they interact with this grid.\footnote{A shape function $S$ must have compact support and satisfy $\sum_g S(\vb{x}_g-\vb{x}_p) = 1$ for any $\vb{x}_p$. \texttt{dHybridR} uses the triangular-shaped cloud (TSC) shape function, whose compact support spans 3 grid points per dimension. These grid points are called the TSC stencil \cite{Hockney1988}.} For macroparticles to move according to their equations of motion, Eq.~\eqref{eq:Lorentz-force-law}, they require the electromagnetic fields at their location $\vb{x}_p$; interpolation from the background grid to the macroparticle location is known as a \textit{gather}, defined by

\begin{equation}
\begin{aligned}
    \vb{E}(\vb{x}_p)
        &= \sum_g \vb{E}(\vb{x}_g) S(\vb{x}_g-\vb{x}_p), \\
    \vb{B}(\vb{x}_p)
        &= \sum_g \vb{B}(\vb{x}_g) S(\vb{x}_g-\vb{x}_p),
\end{aligned}
\label{eq:gather}
\end{equation}

\noindent with the summation over all grid points. Similarly, macroparticles are initialized with a specific \textit{weight} $Q_p$ and \textit{deposit} this weight to the grid points around them, yielding the moments that enter Ohm's law,

\begin{equation}
\begin{aligned}
    \rho_\calK(\vb{x}_g)
        &= \sum_p Q_p S(\vb{x}_g-\vb{x}_p), \\
    \vb{J}_\calK(\vb{x}_g)
        &= \sum_p Q_p\vb{v}_p S(\vb{x}_g-\vb{x}_p),
\end{aligned}
\label{eq:deposit}
\end{equation}

\noindent with the summation over all macroparticles. In \texttt{dHybridR}, a macroparticle's weight is specified at initialization or injection as

\begin{equation}
    Q_p = \frac{q_s n_s(\vb{x}_p)}{N_{\mathrm{ppc},s}}\pqty{1+\frac{\vb{u}_s(\vb{x}_p)\vdot\vb{v}'_p}{c^2}},
\end{equation}

\noindent where $N_{\mathrm{ppc},s}$ is the number of macroparticles of species $s$ instantiated per cell (particles per cell or ppc), $n_s$ and $\vb{u}_s$ are the simulation-frame initial number density and bulk velocity, and $\vb{v}'_p$ is the macroparticle's initial velocity in the species bulk-rest frame. The relativistic correction in parentheses accounts for the relativity of simultaneity between the species bulk-rest frame, in which $\vb{v}'_p$ is sampled, and the simulation frame. One may alternatively account for this effect at the sampling stage rather than in $Q_p$ \cite{Melzani+2013, Zenitani2015}, with the resulting distribution moments equivalent up to sampling noise. A benefit of correcting $Q_p$ is that it leaves sampling unchanged and places no restriction on either the bulk-rest frame distribution or the boost direction.

A timestep begins by sorting macroparticles by cell and depositing them onto the background grid via Eq.~\eqref{eq:deposit}, yielding $\rho_\calK^{n+1/2}$ and $\vb{J}_\calK^{n+1/2}$. The electromagnetic fields are then evolved via a two-step predictor-corrector scheme. The predictor step solves for the pair $\pqty{\vb{B}^{n+1/2}, \vb{E}^{n+1/2}}$ using Faraday's law,

\begin{equation}\label{eq:Faradays-law}
    \pdv{\vb{B}}{t} = -c \curl{\vb{E}},
\end{equation}

\noindent and Ohm's law, Eq.~\eqref{eq:Ohms-law}, via fixed-point iteration while holding the moments $\pqty{\rho_\calK^{n+1/2}, \vb{J}_\calK^{n+1/2}}$ fixed. Specifically, Eq.~\eqref{eq:Faradays-law} is integrated over a half-step using the trapezoid rule,

\begin{equation}
    \vb{B}_{(\ell+1)}^{n+1/2}
        = \vb{B}^n-\frac{c\Delta t}{2} \curl{\pqty{\frac{
           \vb{E}^n+\vb{E}_{(\ell)}^{n+1/2}}{2}}},
\end{equation}

\noindent with the implicit half-step electric field, $\vb{E}_{(\ell+1)}^{n+1/2}$, computed via 

\begin{equation}
    \vb{E}_{(\ell+1)}^{n+1/2} = \operatorname{Ohm}\bqty{\rho_\calK^{n+1/2},\vb{J}_\calK^{n+1/2},\vb{B}_{(\ell+1)}^{n+1/2}},
\end{equation}

\noindent where $\operatorname{Ohm}\bqty{\cdots}$ is the right-hand side of Eq.~\eqref{eq:Ohms-law} for a polytropic $P_e$ and $\ell \in \qty{0, 1, \ldots, N_\ell-1}$. The fixed-point iteration is seeded with the estimate $\vb{E}_{(0)}^{n+1/2} = \vb{E}^n$ and executed until the electric field converges or until a user-specified $N_\ell$ iterations are performed. 

Given $\ell^\star$ iterations performed during the predictor step, the corrector step obtains the final magnetic field through integration of Faraday's law over a full timestep using the midpoint rule,

\begin{equation} \label{eq:corrector}
    \vb{B}^{n+1} = \vb{B}^n - c \Delta t \curl{\vb{E}^{n+1/2}_{(\ell^\star)}}.
\end{equation}

What remains is the mutually consistent advance of $\vb{E}$ from $t^{n}$ to $t^{n+1}$ and the macroparticle state $\pqty{\vb{x}_p,\vb{v}_{p}}$ from $t^{n+1/2}$ to $t^{n+3/2}$. This has long been difficult for hybrid models due to its circular nature: the equation describing the electric field at $t^{n+1}$ is the instantaneous Ohm's law, Eq.~\eqref{eq:Ohms-law}, which is given in terms of the moments $\rho_\calK$ and $\vb{J}_\calK$ at $t^{n+1}$, which are, in turn, given by the macroparticle state at $t^{n+1}$, the midpoint of the push from $t^{n+1/2}$ to $t^{n+3/2}$ that uses $\vb{E}^{n+1}$ itself. Several schemes have been developed to address this issue (see e.g. Refs.~\cite{Winske+2023, Harned1982, Kunz+2014, Stanier+2019}). 

\texttt{dHybridR} addresses this circularity with a second fixed-point iteration over Ohm's law and the macroparticle push while holding the grid $\vb{B}^{n+1}$, Eq.~\eqref{eq:corrector}, fixed. For each iteration $m \in \qty{0, 1, \ldots, N_m-1}$, the electromagnetic fields are gathered at the estimated midpoint position

\begin{equation}\label{eq:midpoint-estimate}
    \vb{x}^{n+1}_{p,(m)} = \vb{x}_p^{n+1/2} + \frac{\Delta t}{2}\vb{v}^{n+1}_{p,(m)},
\end{equation}

\noindent using the grid $\vb{B}^{n+1}$ and a grid estimate $\vb{E}^{n+1}_{g,(m)}$ according to Eq.~\eqref{eq:gather}. Then, the velocity is advanced a full timestep via

\begin{equation}\label{eq:Boris-velocity-update}
    \vb{v}_{p,(m+1)}^{n+3/2} = \operatorname{Boris}\bqty{\vb{v}_p^{n+1/2}, \vb{E}_{p,(m)}^{n+1}, \vb{B}_{p,(m)}^{n+1}},
\end{equation}

\noindent where $\operatorname{Boris}\bqty{\cdots}$ is the standard relativistic Boris pusher \cite{Boris1970, Ripperda+2018} and $\vb{X}_p \equiv \vb{X}(\vb{x}_p)$ for $\vb{X}\in\qty{\vb{E},\vb{B}}$. The macroparticle midpoint state is subsequently computed as

\begin{align}
    \vb{v}_{p,(m+1)}^{n+1}
        &= \frac{1}{2}\left(
           \vb{v}_p^{n+1/2}+\vb{v}_{p,(m+1)}^{n+3/2}\right), \\
    \vb{x}_{p,(m+1)}^{n+1}
        &= \vb{x}_p^{n+1/2}
           +\frac{\Delta t}{2}\vb{v}_{p,(m+1)}^{n+1}. \label{eq:position-estimate}
\end{align}

This midpoint state is then deposited to the background grid via Eq.~\eqref{eq:deposit}, yielding new moments used to compute the next electric field estimate via Eq.~\eqref{eq:Ohms-law},

\begin{equation}
    \vb{E}_{(m+1)}^{n+1} = \operatorname{Ohm}\bqty{\rho_{\calK,(m+1)}^{n+1},\vb{J}_{\calK,(m+1)}^{n+1}, \vb{B}^{n+1}}.
\end{equation}

\noindent The fixed-point iteration is seeded with

\begin{align}
    \vb{E}_{(0)}^{n+1} &= 2\vb{E}^{n+1/2}_{(\ell^\star)} - \vb{E}^n,\\
    \vb{v}^{n+1}_{p,(0)} &= \vb{v}^{n+1/2}_p
\end{align}

\noindent and iterated until the electric field converges or until a user-specified $N_m$ iterations are performed.

After the second fixed-point iteration, with $m^\star$ iterations performed, the final macroparticle state is 

\begin{align}
    \vb{v}_p^{n+3/2} &= \vb{v}^{n+3/2}_{p,(m^\star)},\\
    \vb{x}_p^{n+3/2} &= \vb{x}_p^{n+1/2} + \Delta t\,\vb{v}^{n+1}_{p,(m^\star)}. \label{eq:final-position}
\end{align}

Boundary conditions are then applied to the macroparticles, followed by inter-rank particle exchange and, if enabled, particle injection and load balancing.

\texttt{dHybridR} carries all grid quantities (fields and moments) on two grids $\mathcal{G}_1$ and $\mathcal{G}_2$ offset from one another by half a cell in every dimension, with components collocated at each grid point rather than placed on the faces and edges of a Yee mesh \cite{Yee1966}. Particle-grid operations such as gathers and deposition use $\mathcal{G}_1$. Every derivative operator maps $\mathcal{G}_1$ onto $\mathcal{G}_2$ or vice versa, and since the discrete approximations to $\partial_x, \partial_y$, and $\partial_z$ used to construct the curl commute, the corresponding discrete $\div{\curl{(\cdot)}} = 0$ identically. Advancing the magnetic field with Eq.~\eqref{eq:corrector} then conserves the discrete $\div{\vb{B}}$ exactly. The hybrid code \texttt{Pegasus} instead enforces this condition using constrained transport \cite{Kunz+2014}.

\begin{table}[pos=t]
\caption{Memory hierarchy for the Intel Xeon Max 9470C CPUs on an Aurora compute node. Capacity is per core for the registers, L1, and L2 caches; per socket for the L3 cache; and per node for main memory. Register capacity is for the AVX-512 vector registers only, which comprise their majority.}
\label{tab:memhier}
\centering
\footnotesize
\begin{tabular}{@{}llr@{}}
\toprule
Level & Location & Capacity \\
\midrule
Registers   & in the core, on-chip & 2 KiB   \\
L1 cache    & per core, on-chip    & 48 KiB  \\
L2 cache    & per core, on-chip    & 2 MiB   \\
L3 cache    & shared, on-chip      & 105 MiB \\
Main memory &                      &         \\
\quad DDR5  & on the motherboard   & 1 TiB   \\
\quad HBM2e & in the CPU package   & 128 GiB \\
\bottomrule
\end{tabular}
\end{table}

\section{CPU-side groundwork}\label{sec:cpu}

Before being ported to GPUs, the codebase was refactored to take advantage of the memory hierarchy and vector instructions ubiquitous in modern CPUs. Appendix~\ref{app:cpu-refactor} reviews these concepts for an Aurora\footnote{Aurora is a U.S. Department of Energy (DOE) exascale supercomputer at Argonne National Laboratory.} compute node (Table~\ref{tab:memhier}) and describes \texttt{dHybridR}'s array layouts before and after the refactor. In sum, the species property arrays and the electromagnetic field arrays were reorganized so that consecutive iterations of a loop over macroparticles or grid cells read adjacent elements in memory (unit stride accesses). This makes efficient use of the cache hierarchy and allows the compiler to auto-vectorize such loops, i.e. to employ vector instructions that operate on multiple values at once, a form of data-level parallelism known as \textit{single instruction, multiple data} (SIMD).

Particle sorting has long been known to be an important aspect of PIC code optimization \cite{Bowers2001}. Without it, particles that are initially spatially close do not remain so as the simulation progresses; the gather and deposit operations, Eqs.~\eqref{eq:gather} and \eqref{eq:deposit}, then involve electromagnetic field and moment read and write operations at grid points that are scattered across main memory. Given that their distance in memory typically exceeds a cache line width, most of these accesses result in cache misses that degrade performance. This is especially important on modern supercomputers, where the maximum floating-point performance has grown faster than memory bandwidth. For example, the CPU of Table~\ref{tab:memhier} has a single-precision main-memory (DDR5) ridge point of about $24$ FLOPs/byte in the roofline model \cite{Williams+2009}. If the moment and field stencil data are repeatedly fetched from main memory, the gather and deposit operations have a main-memory arithmetic intensity on the order of $0.1$ FLOPs/byte, limiting their performance to around $0.5\%$ of its theoretical maximum. With field and moment data retrieved from cache, the main-memory arithmetic intensity rises by more than an order of magnitude, allowing the gather and deposit to move closer toward peak machine floating-point performance.

Given this, we modified \texttt{dHybridR} to sort macroparticles every step rather than periodically. The properties of each cell's macroparticles are then stored contiguously in memory, enabling the gather, push, and deposit routines to loop over grid cells and, within each cell, over its macroparticles. In the gather step, the $3^D$ electromagnetic field grid points for spatial dimension $D\in\qty{1,2,3}$ within the compact support of the TSC shape function are the same for every macroparticle whose midpoint estimate remains within the cell, so they are read once and remain in cache or registers as we loop over the macroparticles in that cell. In the deposit step, we similarly accumulate each macroparticle's contribution to the moments in small cache-resident arrays, updating the global grid only once per cell.  Additionally, within each cell the loop over macroparticles reads every property array with unit stride, enabling auto-vectorization by the compiler. \texttt{Smilei} similarly uses cell-based particle sorting every timestep to support vectorization in its gather and deposition routines \cite{Beck+2019}. With this choice, sorting takes roughly a third of the much shorter refactored timestep compared with an amortized fraction of a few percent in the original's.

\begin{figure}[pos=t]
\centering
\includegraphics[width=\columnwidth]{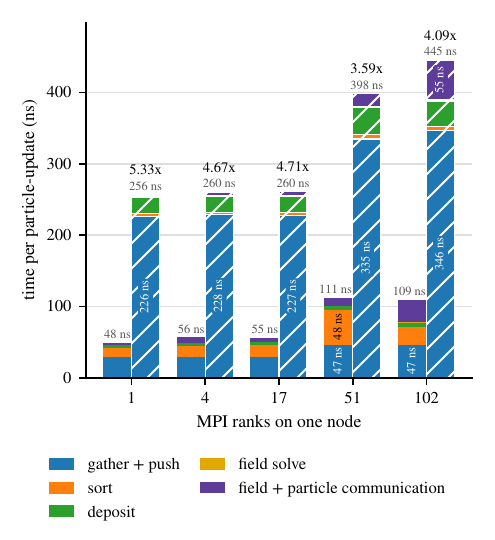}
\caption{\label{fig:timer-breakdown}Wall-clock time required to update one macroparticle in each part of the \texttt{dHybridR} timestep loop for the refactored (solid) and original (hatched) CPU codes using a $2000^2$ cell grid at 256 ppc. Colors identify the different timed routines in a timestep. The original code's every-10-step sort cost is shown as its (1/10)\textsuperscript{th} amortized cost.}
\end{figure}

\begin{figure}[pos=!htb]
\centering
\includegraphics[width=\columnwidth]{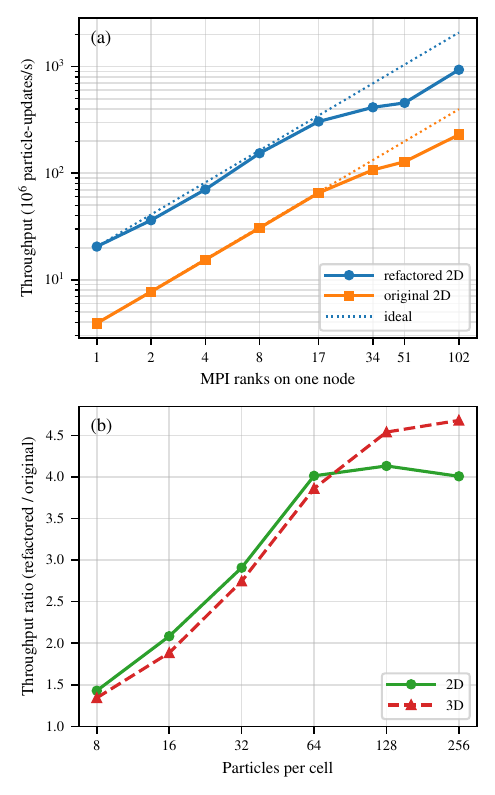}
\caption{\label{fig:cpu-scaling}CPU performance before and after refactoring on one Aurora node. (a) Throughput $F(R)$ for a $2000^2$ grid at $256$ ppc as a function of rank count. (b) Refactored-to-original throughput ratio vs. ppc using 102 ranks for $2000^2$ (2D) and $200\times 200 \times 100$ (3D) cell grids.}
\end{figure}

To quantify performance, \texttt{dHybridR} uses a figure of merit (FOM) or ``throughput'' of particle-updates per second, where $X$ ``particle-updates'' means that $X$ macroparticles were evolved over an entire timestep. It is computed over windows of 10 consecutive timesteps to avoid every-step MPI reduction overhead, and given by

\begin{equation}\label{eq:FOM}
    F(R) = \frac{1}{T_{\rm max}}\sum_{r=1}^R\sum_{j=1}^{10} N_{r,j},
\end{equation}

\noindent where $R$ is the total number of MPI ranks, $N_{r,j}$ is the number of particles updated by rank $r$ in timestep $j$ out of 10, and $T_{\rm max} = \max_r \qty{t_{{\rm end},r} - t_{{\rm start},r}}$ is the time taken by the slowest rank over those 10 steps. Throughputs measured over longer intervals are the total particle-updates divided by the total elapsed time (including any output, load balancing, or particle injection), i.e. the duration-weighted average $\overline{F}(R) = \sum_w T_{{\rm max},w} F_w(R) / \sum_w T_{{\rm max},w}$ of the values $F_w(R)$ of Eq.~\eqref{eq:FOM} over the individual 10-step windows $w$ in that interval.

The benchmark used here and in Sec.~\ref{sec:performance} is a uniform thermal plasma containing a single proton species with periodic boundary conditions. Each component of a macroparticle's velocity is independently sampled from a zero-mean Gaussian distribution with standard deviation $v_{\rm th}=v_{A0}$, where $v_{A0}=d_i \Omega_i$ is the reference Alfv\'en speed, $d_i = c/\omega_{pi}$ and $\Omega_i = e B_0/(m_p c)$ are the ion (proton) inertial length and cyclotron frequency at the reference density $n_0$ and magnetic field $B_0$, and $\omega_{pi} = \sqrt{4\pi n_0 e^2/m_p}$. Simulations are run in single precision with a uniform cell width of $0.5\,d_i$ and a timestep of $\Delta t\, \Omega_i = 10^{-3}$. The initial electromagnetic fields are zero and electrons are taken to obey a $\gamma_e = 5/3$ polytropic equation of state. Furthermore, we use $c=100\,v_{A0}$, perform two iterations of each fixed-point solve, and disable output, load balancing, and particle injection. All CPU runs bind one MPI rank to one physical CPU core, regardless of whether the processor supports multiple logical cores via simultaneous multithreading (SMT). We find that SMT yields no performance benefit or negatively affects performance. Aurora nodes, in particular, have both DDR5 and high-bandwidth (HBM2e) main memory, the latter providing significantly more memory bandwidth than the former. We therefore use the \texttt{numactl -{}-preferred} flag for each MPI process to favor HBM, only falling back to DDR when necessary. 

Figure~\ref{fig:timer-breakdown} shows the time required for a rank to process a macroparticle in each part of a \texttt{dHybridR} timestep for the original and refactored CPU implementations, computed as $t_{{\rm pu},k} = \avg{T_k} / \avg{N}$ where $T_k(r)$ is the accumulated wall-clock time for rank $r$ measured for part $k$ of the timestep, $N(r)$ the number of particle-updates performed by rank $r$ over the same measurement interval, and $\avg{\cdot}$ the rank average. Solid bars denote the refactored code and hatched bars the original code. Measurements were performed on a single Aurora node using a $2000^2$ cell grid with 256 ppc.

The gather/push and deposit routines are $7.8\times$ and $6.9\times$ faster at one rank, and $7.4\times$ and $6.4\times$ faster at 102 ranks, respectively. Individual sorts are also faster, but sorting every timestep rather than every tenth timestep makes the average sorting cost per timestep up to $7.2\times$ as large across ranks. The overall timestep yields a speedup of approximately $5\times$ on one core and $4\times$ on the full Aurora node, with Aurora's CPU-attached HBM particularly benefiting the refactored code's sorting and deposition routines. 

Figure~\ref{fig:cpu-scaling}(a) shows the throughput for both codes in 2D as we move from a single rank to the entire Aurora node. The refactored code reaches $2.05\times 10^7$ particle-updates/s on a single core and $9.35\times 10^8$ on a full node in 2D compared with $3.91\times 10^6$ and $2.30\times 10^8$ particle-updates/s for the original. Both codes scale well at small rank counts, but fall short of ideal strong scaling in large part due to increases in gather/push and communication costs (and sorting costs for the refactored code).

Figure~\ref{fig:cpu-scaling}(b) shows the refactored-to-original throughput ratio as a function of ppc for a full Aurora node. The ratio rises from about $1.4$ at 8 ppc to $4.0$ in 2D and $4.7$ in 3D at 256 ppc. This can be interpreted using a simple cost model. For a fixed grid size, we may approximate the wall-clock time elapsed during one timestep as $T(N_{\rm ppc}) \simeq \alpha + \beta N_{\rm ppc}$, where $N_{\rm ppc}$ is the ppc, $\alpha$ represents ppc-independent time, and $\beta$ is time per unit increase in ppc. The speedup of the refactored implementation over the original implementation is then

\begin{equation}\label{eq:ppc-speedup}
\mathcal{S}(N_{\rm ppc}) = \frac{T_{\rm orig}(N_{\rm ppc})}{T_{\rm ref}(N_{\rm ppc})} \simeq
\frac{\alpha_{\rm orig} + \beta_{\rm orig} N_{\rm ppc}}
     {\alpha_{\rm ref} + \beta_{\rm ref} N_{\rm ppc}},
\end{equation}

\noindent which increases with ppc when the ppc-dependent speedup $\beta_{\rm orig}/\beta_{\rm ref}$ exceeds the ppc-independent speedup $\alpha_{\rm orig}/\alpha_{\rm ref}$ and asymptotically approaches $\beta_{\rm orig}/\beta_{\rm ref}$ at large ppc. Both curves in Fig.~\ref{fig:cpu-scaling}(b) are generally consistent with this interpretation, demonstrating increasing throughput ratios that begin to flatten at higher ppc.

\section{GPU implementation}\label{sec:gpu-implementation}

The CPU optimizations of Sec.~\ref{sec:cpu} exploit data-level parallelism and the memory hierarchy, principles equally applicable on GPUs. We had three design goals for the subsequent GPU port: support CPU and GPU execution with a single codebase to avoid maintaining separate versions; avoid dependence on a single GPU vendor, given DOE leadership computing facilities already span Intel (Aurora), AMD (Frontier), and NVIDIA (Polaris) hardware; and retain the ability to write specialized GPU kernels for the most expensive routines, where explicit control over the parallel execution and memory would enable further optimization. We therefore decided to use OpenMP target offload throughout the Fortran codebase. For kernels that were deemed to have potential for significant speedup with a lower-level approach, we wrote GPU-specific SYCL kernels that were connected to the primary Fortran codebase using C interoperability. 

\begin{listing}[tb]
\begin{lstlisting}[language=Fortran,title={Fortran side},captionpos=t]
interface
   subroutine sycl_deposit( &
      x, y, Npart, rho_K, J_K, Ncells &
   ) bind(C)
      import :: c_ptr, c_int
      type(c_ptr), value :: x, y, rho_K, J_K
      integer(c_int), value :: Npart, Ncells
   end subroutine
end interface

!$omp target data use_device_addr(x, y, rho_K, J_K)
call sycl_deposit( &
   c_loc(x), c_loc(y), Npart, &
   c_loc(rho_K), c_loc(J_K), Ncells &
)
!$omp end target data
\end{lstlisting}
\begin{lstlisting}[language=C++,title={C++ side},captionpos=t,firstnumber=last]
static sycl::queue q{
   sycl::gpu_selector_v, 
   sycl::property::queue::in_order()
};

extern "C" void sycl_deposit(
   float* x, float* y, int Npart,
   float* rho_K, float* J_K, int Ncells
) {
   q.submit([&](sycl::handler& h) {
      h.parallel_for(
         sycl::nd_range<1>{global, local},
         [=](sycl::nd_item<1> it) {
            // kernel implementation
         });
   }).wait();
}
\end{lstlisting}
\caption{Simplified representation of the Fortran-to-SYCL interface for charge and current deposition used in \texttt{dHybridR}. All arrays are device-resident throughout a timestep. OpenMP's \texttt{use\_device\_addr} exposes their device addresses, allowing them to be passed to the corresponding SYCL kernel through a \texttt{bind(C)} interface.}
\label{lst:fortran-sycl-interface}
\end{listing}

Listing~\ref{lst:fortran-sycl-interface} gives an example of the Fortran-to-SYCL connection. On the Fortran side, a \texttt{bind(C)} interface to the C++ wrapper \texttt{sycl\_deposit} is declared. The call is enclosed in an OpenMP \texttt{target data} region with \texttt{use\_device\_addr}, so that \texttt{c\_loc} returns the addresses of the device-resident arrays rather than the addresses of their host counterparts. On the C++ side, the wrapper is an \texttt{extern "C"} function that submits the kernel to a SYCL queue. This approach is vendor-agnostic, only requiring support for OpenMP offload, SYCL, and their interoperability on the target GPU architecture.

At program initialization, all data are offloaded from the CPU (the \textit{host}) to the GPU (the \textit{device}) and remain on-device throughout a typical timestep (the exception being particle injection and occasional load balancing, which remain on host) to avoid costly host-device synchronization. Most kernels over grid cells or macroparticles are simply decorated with an OpenMP \texttt{target teams distribute parallel do} directive, allowing them to execute on the device-resident data. We additionally use GPU-aware MPI to avoid device-host-device round-trip synchronizations for inter-rank halo exchanges and particle communication. However, profiling revealed that the moment deposition and fused gather/push kernels would benefit from explicit control of subgroup execution and the memory hierarchy, so they were written in SYCL (Secs.~\ref{sec:deposition} and \ref{sec:fused-gather-push}). We detail these kernels in the following subsections. 

The smallest unit of parallel execution in SYCL is a \textit{work-item} (thread on NVIDIA and AMD). Work-items are grouped into \textit{subgroups} (warps on NVIDIA, wavefronts on AMD) that SIMD-execute. Each work-item within a subgroup is often called a \textit{lane}. Subgroups in turn form \textit{work-groups} (thread blocks on NVIDIA and AMD), where members of a work-group can share programmer-managed \textit{local memory} (shared memory on NVIDIA and AMD) and synchronize their execution through barriers. On Aurora's Ponte Vecchio (PVC) GPUs, each work-group executes within a single Xe-core (analogous to a streaming multiprocessor on NVIDIA or a compute unit on AMD), with multiple work-groups capable of residing on the same Xe-core given sufficient hardware resources.

The number of resident subgroups on an Xe-core relative to the hardware maximum is known as \textit{occupancy}. Subgroups become resident on an Xe-core as entire work-groups, so some capacity may remain unused when another work-group cannot fit. Lower occupancy can reduce the GPU's ability to hide latency by executing a ready subgroup while another waits (for a memory access to complete, for example), potentially degrading performance. Table~\ref{tab:gpu-memory} summarizes the GPU memory hierarchy.

The cell ordering described in Sec.~\ref{sec:cpu} is maintained on the GPU using a device-resident counting sort. The per-cell macroparticle counts and their prefix sum determine where each cell's macroparticles begin and end in the species property arrays. The species arrays are then sorted by copying each macroparticle's properties to a distinct index within its cell's index range, with no prescribed order among macroparticles in the same cell.

\begin{table}[pos=t] 
\caption{Memory hierarchy of Aurora's Intel Data Center GPU Max 1550 (Ponte Vecchio) \cite{Intel-GPU-optimization-guide}. One GPU consists of two PVC tiles. Registers hold work-items' private values, with capacity given per Xe-core. Each Xe-core has an L1 cache, a portion of which is available for allocation as programmer-managed local (shared) memory. The last-level cache is the final cache before global memory, analogous to a CPU's L3 cache. Global-memory data are stored in high-bandwidth memory (HBM2e), which serves as the GPU's main memory.}
\label{tab:gpu-memory}
\centering
\footnotesize
\begin{tabular}{@{}llr@{}}
\toprule
Storage & Scope & Capacity \\
\midrule
Registers
  & Per Xe-core & 512 KiB \\
L1 cache
  & Per Xe-core & 512 KiB \\
\quad Local (shared) memory
  & Per Xe-core & 128 KiB \\
Last-level cache
  & Per tile & 192 MiB \\
Global memory
  & Per tile & 64 GiB \\
\bottomrule
\end{tabular}
\end{table}

\subsection{Charge and current deposition} \label{sec:deposition}

The formulae for charge and current deposition, Eq.~\eqref{eq:deposit}, add charge and current contributions to the $3^D$ grid points of the TSC stencil for each of the four moment components $(\rho_\calK, \vb{J}_\calK)$. Macroparticles in the same cell share this stencil, so an implementation that parallelizes work across macroparticles will have macroparticles trying to update the same moment array elements at the same time. Each update must therefore be an atomic read-modify-write operation. For a cell containing $N_{\rm ppc}$ macroparticles per species, this gives $4\times 3^D N_{\rm ppc}$ atomic updates per species at every deposition kernel call applied to just $4\times 3^D$ distinct array elements. With each element in the TSC stencil receiving $N_{\rm ppc}$ of these atomic updates, they are highly serialized. 

An improved algorithm takes advantage of the limited particle displacement over a timestep. Due to the CFL condition for the potentially speed-of-light kinetic ions, the timestep is bounded above by

\begin{equation}
    \Delta t \leq \frac{C_{\rm max}}{v_{\rm max} \sqrt{\sum_{i=1}^D (1/\Delta x_i)^2}},
\end{equation}

\noindent where $\Delta x_i \in \qty{\Delta x, \Delta y, \Delta z}$, $v_{\rm max} = \max_p \qty{v_p}$ is the maximum macroparticle speed (conservatively taken to be the speed of light $c$), and $C_{\rm max} \in (0,1\rbrack$ is a safety factor. Under this constraint, no macroparticle moves more than the smallest cell width over a timestep, with the fraction of ``crosser'' particles leaving their cell $N_{\rm cross} / N_{\rm ppc} \ll 1$ typically. Since the particles are sorted each step, we can assign each cell a corresponding subgroup to reduce global atomic additions.

\begin{figure*}[pos=t]
\centering
\includegraphics[width=\textwidth]{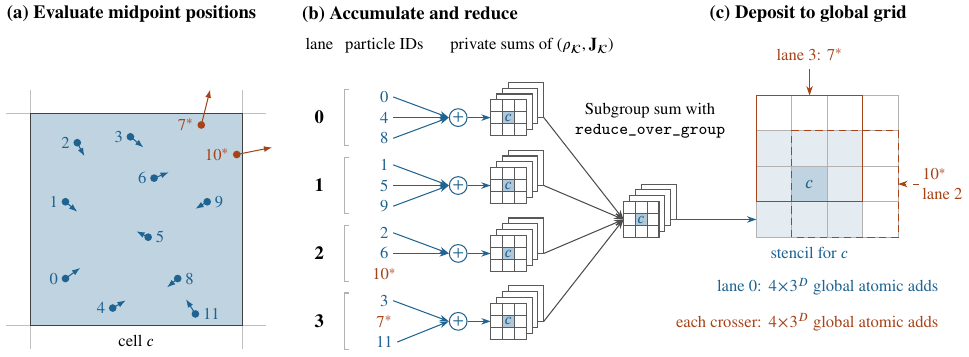}
\caption{\label{fig:subgroup-per-cell-deposit}The GPU charge and current deposition kernel, illustrated in 2D for 12 macroparticles of a single species and a subgroup of 4 lanes, one subgroup per cell. (a) Arrows connect the initial position of each macroparticle to its estimated midpoint location. (b) Each lane accumulates contributions to charge and current for the non-crossing particles into four private $3^2$ stencils. A subgroup reduction combines stencils element-wise across lanes. (c) Lane 0 atomically deposits the reduced stencils to the large moment arrays residing in global memory, shown in blue. Crossing particles atomically deposit their individual shifted stencils as well (orange).}
\end{figure*}

Our improved deposition routine is shown in Fig.~\ref{fig:subgroup-per-cell-deposit}. At the beginning of the deposition step in the second fixed-point iteration, a macroparticle's temporary position estimate is given by Eq.~\eqref{eq:position-estimate}. The CFL condition limits the displacement to one cell, so macroparticles that remain within their original cell share the same $3^D$ TSC stencil.

Each lane of the subgroup accumulates its non-crossing macroparticle contributions into four private $3^D$ arrays, one for $\rho_\calK$ and one for each current component, $J_{\calK,x}, J_{\calK,y},$ and $J_{\calK,z}$. Once complete, the lanes' private arrays are summed element-wise across the subgroup with SYCL's \texttt{reduce\_over\_group} collective. Lane~0 then writes the reduced sums to the moment arrays in global memory using $4\times 3^D$ additions, which must be atomic because the neighboring cells' stencils overlap. Macroparticles with estimated midpoint positions that lie outside their original cell instead deposit directly to their shifted stencils through their assigned lanes, using $4\times 3^D$ global atomic additions per crossing macroparticle.

This improved routine requires $4\times 3^D (1+N_{\rm cross})$ global atomic additions compared with $4\times 3^D N_{\rm ppc}$ in the naive implementation. By accumulating the depositions in private registers and reducing over each cell's subgroup, the number of atomic read-modify-write operations scales with the number of cells plus the total number of crossing macroparticles rather than the total number of macroparticles. For $N_{\rm cross}/N_{\rm ppc}\ll 1$, this reduces both the total number of global memory accesses and the number of them directed at any single array element, the latter addressing the serialization described above. The remaining atomic updates issued by lane~0 of each subgroup and by crossing macroparticles are few enough in number to avoid limiting the resulting kernel. 

\subsection{Field gather and particle push} \label{sec:fused-gather-push}

Each iteration of the second fixed-point solve requires a gather of the electromagnetic fields at the estimated midpoint position, Eq.~\eqref{eq:midpoint-estimate}, and a subsequent Boris velocity update, Eq.~\eqref{eq:Boris-velocity-update}. \texttt{dHybridR} fuses these operations into a single SYCL kernel so that the gathered electromagnetic fields are immediately consumed by the Boris push, avoiding intermediate writes to and reads from global memory.

The kernel execution parallels the deposition routine: subgroups and cells are in one-to-one correspondence and employ the same stay/cross split. Macroparticles with estimated midpoint positions within their original cell share the same $3^D$ grid points for $\vb{E}$ and $\vb{B}$; each subgroup therefore begins by cooperatively loading the corresponding $6\times 3^D$ field values into local (shared) memory. After a work-group barrier, each lane then performs the gather operation, Eq.~\eqref{eq:gather}, for each of its assigned macroparticles using the macroparticle's individual shape-function weights. Macroparticles with midpoint estimates beyond the cell boundary perform the gather operation with shifted electromagnetic field stencils read directly from global memory. Because each lane independently computes the interpolated electromagnetic fields $\pqty{\vb{E}_p, \vb{B}_p}$ for each of its assigned macroparticles, the gather operation requires neither atomic additions nor subgroup reductions across lanes.

Like the authors of OSIRIS, we found that caching electromagnetic field values in local memory improved performance \cite{Lee+2025}. A natural alternative to the above implementation would be to remove the stay/cross split entirely, instead caching $6\times 5^D$ field values per subgroup in local memory that all macroparticles could use. However, our three-dimensional Aurora tests found that this expanded use of local memory made the fused gather/push kernel $2\times$ to $4\times$ slower with a larger penalty at lower ppc, primarily due to lower occupancy and increased field-loading work. The occupancy penalty can be understood using, for example, our Aurora kernel launch configuration: our work-groups contain 256 work-items arranged into 16-lane subgroups, giving 16 subgroups per work-group. In the regular register mode used to execute the kernel, each subgroup holds 128 registers of 64 bytes each, totaling 8~KiB, so the 512~KiB register file of Table~\ref{tab:gpu-memory} holds at most 64 subgroups (four work-groups) per Xe-core \cite{Intel-GPU-optimization-guide}. In three dimensions and single precision, the compact and expanded caches request $\simeq 10\,\textrm{KiB}$ and $\simeq 47\,\textrm{KiB}$ of local memory per work-group\footnote{Specifically, $(16\ \text{subgroups}) \times (6w^{3}\ \text{values/subgroup}) \times (4\ \text{bytes/value})$ for cache widths of $w = 3$ or $5$.}, which the GPU rounds up to its nearest allocation size, 16~KiB and 48~KiB respectively \cite{Intel-GPU-occupancy-guide}. Within the Xe-core's 128~KiB local memory capacity (Table~\ref{tab:gpu-memory}), all four compact-cache work-groups fit, but only two expanded-cache work-groups fit, halving maximum occupancy. Additionally, cooperative loading of the field values into local memory requires 750 field values per cell instead of 162 for the more compact cache. In our tests, these costs outweigh the benefit of a larger local memory stencil.

\subsection{Particle exchange and compaction} \label{sec:particle-exchange-compaction}

\begin{figure}[pos=t]
\centering
\includegraphics[width=\columnwidth]{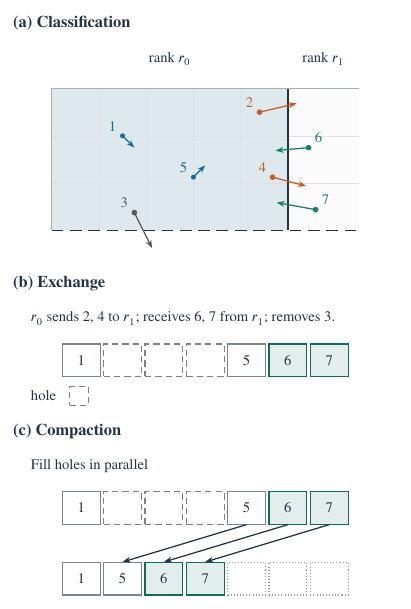}
\caption{\label{fig:particle-exchange}Illustration of parallel particle-exchange and compaction for rank $r_0$, with numbers identifying unique macroparticles throughout. (a) Each rank classifies macroparticles as remaining local to the rank (blue), outgoing to another rank (orange), or leaving through an open boundary (gray). Incoming macroparticles from rank $r_1$ are shown in green. (b) Outgoing macroparticles are exchanged with neighboring ranks and received macroparticles are appended to local arrays. (c) Macroparticles are copied into holes in parallel, and the total macroparticle count adjusted.}
\end{figure}

After the second fixed-point solve, macroparticle positions are advanced to $t^{n+3/2}$ using Eq.~\eqref{eq:final-position}. Macroparticles that cross into a different rank must be sent to that rank, and macroparticles that pass through an open boundary must be eliminated. A macroparticle eliminated from the simulation or sent to a neighboring rank leaves an unused index or \textit{hole} in its species' rank-local property arrays (e.g. position \texttt{x}). Since loops over macroparticles assume that every index of their property arrays corresponds to a valid macroparticle, these holes must be removed via a compaction routine.

The original CPU implementation of compaction was inherently serial: for example, an array \texttt{[A,$\cdot$,C,D,$\cdot$,F]} with two holes would first be compacted to \texttt{[A,$\cdot$,C,D,F]}, then \texttt{[A,F,C,D]}. Here, F is copied from index 6 to 5 and then 5 to 2; since the second copy reads an entry written by the first, it precludes parallelized execution. We therefore refactored the inter-rank particle exchange and compaction routine into a parallelizable three-step procedure that can execute on-device, shown in Fig.~\ref{fig:particle-exchange}. 

The first step, classification, is performed in parallel over macroparticles, using each macroparticle's updated position to determine whether it stays rank-local, must be sent to a neighboring rank, or must be eliminated if it crosses an open boundary. The indices of particles that must be sent or eliminated are only recorded, with no array modifications occurring at this stage. 

In the second step, exchange, outgoing macroparticles are packed into send buffers in parallel and exchanged using GPU-aware MPI. Macroparticles received from neighboring ranks are then appended to the end of the species arrays. 

Finally, during compaction, the number of surviving macroparticles determines the last array index to retain. Holes at or below this index are paired one-to-one with surviving macroparticles at higher indices, allowing their properties to be copied into the holes in parallel. The total macroparticle count is then adjusted.

\section{Performance}\label{sec:performance}

\begin{table*}[pos=t]
\centering
\caption{Top block: Machines used in the benchmarks of Sec.~\ref{sec:performance}. Aurora reserves the first CPU core in each socket for system resources, while Frontier reserves the first core per L3 cache region; available physical cores are given in parentheses. Bottom block: Cell counts, global grid sizes, and ppc for each benchmark of Sec.~\ref{sec:performance}. Total ranks $R = R_x R_y$ in 2D and $R_x R_y R_z$ in 3D, where $R_x, R_y, R_z$ are the number of ranks along each axis. MPI ranks per node are for GPU runs.}
\label{tab:perf-methods}
\small

\begin{tabular*}{\textwidth}{@{\extracolsep{\fill}}lrlllr@{}}
\toprule
System & Nodes & Host CPUs & GPUs/node & Accelerators/GPU & MPI ranks/node \\
\midrule
Aurora & 10,624
  & $2\times$52-core Intel Xeon Max (51 usable)
  & $6\times$Intel GPU Max 1550 & 2 tiles & 12 \\
Frontier & 9,856
  & $1\times$64-core AMD EPYC (56 usable)
  & $4\times$AMD MI250X & 2 GCDs & 8 \\
Polaris & 560
  & $1\times$32-core AMD EPYC
  & $4\times$NVIDIA A100 & 1 GPU & 4 \\
\bottomrule
\end{tabular*}

\par\medskip

\begin{tabular*}{\textwidth}{@{\extracolsep{\fill}}llllr@{}}
\toprule
Benchmark & Cells & Global 2D grid & Global 3D grid & ppc \\
\midrule
\multicolumn{5}{@{}l}{Sec.~\ref{sec:perf-strong-weak-scaling}: Strong/weak scaling} \\
\quad Strong scaling
  & $4\times 10^6$ & $2000\times2000$ & $200\times200\times100$ & 256 \\
\quad Weak scaling
  & $R\times 10^6$ & $1000R_x\times1000R_y$
  & $100R_x\times100R_y\times100R_z$ & 256 \\
\multicolumn{5}{@{}l}{Sec.~\ref{sec:perf-cpu}: CPU/GPU comparison} \\
\quad Isolated accelerator/core/socket
  & $1\times 10^6$ & $1000\times1000$ & $100\times100\times100$ & Varied \\
\quad Full node
  & $4\times 10^6$ & $2000\times2000$ & $200\times200\times100$ & Varied \\
\multicolumn{5}{@{}l}{Sec.~\ref{sec:perf-energy}: Energy efficiency} \\
\quad Full node
  & $4\times 10^6$ & $2000\times2000$ & $200\times200\times100$ & Varied \\
\bottomrule
\end{tabular*}
\end{table*}

We assess the performance of GPU-accelerated \texttt{dHybridR} on Aurora, Frontier, and Polaris\footnote{Frontier is a U.S. Department of Energy exascale supercomputer at Oak Ridge National Laboratory, and Polaris is a smaller cluster at Argonne National Laboratory.} (top block of Table~\ref{tab:perf-methods}) through strong and weak scaling studies, comparison between the GPU and refactored CPU implementations, and energy efficiency measurements. All benchmarks use the thermal-plasma setup of Sec.~\ref{sec:cpu} on the grids of the bottom block. We report performance as the FOM of Eq.~\eqref{eq:FOM} and quantify strong and weak scaling with the parallel efficiency

\begin{equation}
    \mathcal{E}(R) = \frac{F(R)/R}{F(R_0)/R_0},
\end{equation}

\noindent where $F(R)$ is the figure of merit of Eq.~\eqref{eq:FOM} and $R_0$ is the total number of MPI ranks used in the machine-specific baseline simulation. 

\texttt{dHybridR} decomposes the simulation domain into $R = R_x \times R_y \times R_z$ blocks with one MPI rank per block. GPU runs use one rank per accelerator, where an accelerator is one GPU tile on Aurora, one graphics compute die (GCD) on Frontier, and one A100 GPU on Polaris (Table~\ref{tab:perf-methods}). 

\subsection{Strong and weak scaling} \label{sec:perf-strong-weak-scaling}

\begin{figure}[pos=t]
\centering
\includegraphics[width=\columnwidth]{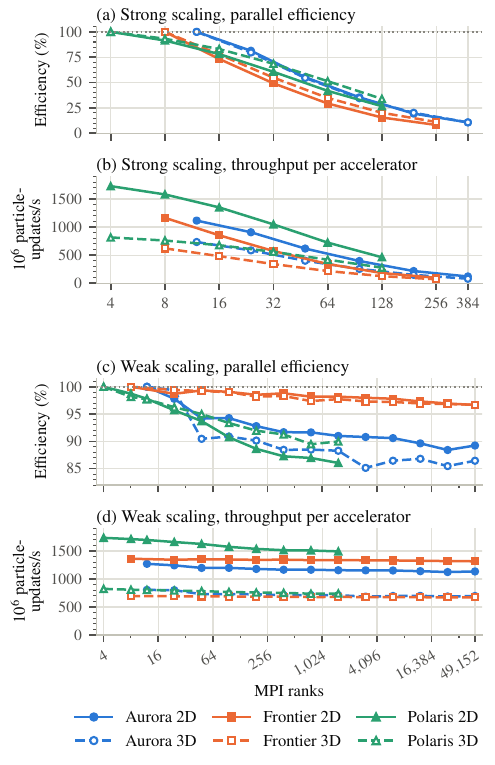}
\caption{\label{fig:strong-weak-scaling}Strong and weak scaling of the GPU-accelerated \texttt{dHybridR} on Aurora, Frontier, and Polaris in 2D and 3D. In subplots (a) and (c), each curve is normalized to its one-node baseline with dotted horizontal lines denoting ideal scaling. Subplots (b) and (d) show the same results in terms of the throughput per accelerator.}
\end{figure}

Figure~\ref{fig:strong-weak-scaling} shows strong (a, b) and weak (c, d) scaling across Aurora, Frontier, and Polaris, in terms of parallel efficiency and the corresponding throughput per accelerator. With an increasing number of ranks, the strong scaling parallel efficiency drops because, although particle-related routines drop in cost, field communication routine costs remain roughly constant. In contrast, weak scaling efficiency remains high on all three machines because each step's MPI communication consists of particle and halo exchanges between neighboring ranks plus a few scalar \texttt{MPI\_Allreduce} collectives, so the amount of communication per rank is somewhat independent of rank count. For each machine, the largest runs correspond to 49,152 ranks on 4,096 Aurora nodes, the same number of ranks on 6,144 Frontier nodes, and 1,536 ranks on 384 Polaris nodes. Weak efficiency remains at $89\%$ (2D) and $86\%$ (3D) on Aurora, $97\%$ (2D and 3D) on Frontier, and $86\%$ (2D) and $90\%$ (3D) on Polaris. In terms of throughput per accelerator (Fig.~\ref{fig:strong-weak-scaling}(d)), these efficiencies correspond to $1100\times 10^{6}$, $1300\times 10^{6}$, and $1500\times 10^{6}$ particle-updates/s on Aurora, Frontier, and Polaris in 2D and roughly $700\times 10^6$ particle-updates/s on all three machines in 3D. On Aurora, the majority of the weak efficiency loss occurs over the first four nodes. These efficiencies compare well with full PIC GPU codes run at similar scale \cite{Fedeli+2022, Myers+2021, Bird+2022}, while the hybrid PIC GPU codes of Sec.~\ref{sec:intro} have no reported scaling beyond 64 GPUs.

\subsection{Speedup relative to the CPU implementation} \label{sec:perf-cpu}

\begin{figure}[pos=t]
\centering
\includegraphics[width=\columnwidth]{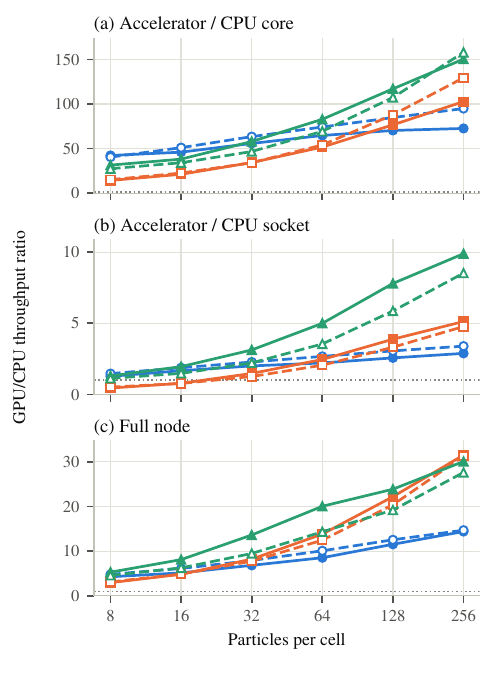}
\caption{\label{fig:cpu-gpu-speedup}GPU-to-CPU throughput ratio vs. ppc for (a) one accelerator compared to one CPU core, (b) one accelerator compared to one CPU socket, and (c) a full node's accelerators compared to its CPUs. The legend is identical to Fig.~\ref{fig:strong-weak-scaling}. Dotted horizontal lines denote a throughput ratio of unity.}
\end{figure}

To quantify the benefit of GPU acceleration, we provide three different comparisons with the refactored CPU code of Sec.~\ref{sec:cpu}. First, comparing one accelerator to one CPU core establishes a reference point independent of how many cores are present in the CPU; comparing an accelerator against an entire CPU socket is a fairer benchmark, taking into account the CPU's multicore parallelism; and comparing a full node's accelerators against its CPUs quantifies how much more science can be achieved given a certain allocation of node-hours. Each comparison is provided as a GPU-to-CPU throughput ratio (speedup) in Fig.~\ref{fig:cpu-gpu-speedup} as a function of ppc. 

All three comparisons vary dramatically with the ppc, with larger ppc providing larger speedups. Aurora's lower throughput ratios at high ppc partly reflect its CPU-attached HBM2e memory, not present on Frontier or Polaris, which increases the refactored CPU code's full-node throughput by approximately $30\%$ at 256 ppc relative to DDR5. 

These curves are again interpretable and generally consistent with the cost model of Sec.~\ref{sec:cpu}. However, unlike the CPU ratio of Fig.~\ref{fig:cpu-scaling}(b), almost none of the curves have approached their asymptotic throughput ratio by 256 ppc, so one would expect continued speedup at larger ppc.

\subsection{Energy efficiency} \label{sec:perf-energy}

\begin{figure}[pos=t]
\centering
\includegraphics[width=\columnwidth]{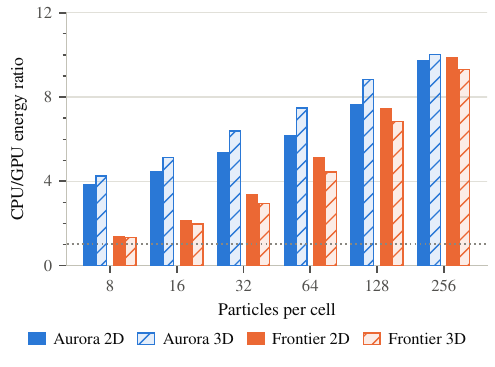}
\caption{\label{fig:node-energy-gain-combined}Ratio of the CPU energy per particle-update to its GPU counterpart, using a full node's accelerators and CPUs. The dotted horizontal line denotes an energy ratio of unity.}
\end{figure}

For the full-node comparisons of Fig.~\ref{fig:cpu-gpu-speedup}(c), we sampled the cumulative energy counters provided by the node, $E(t)$, at $1\,\text{Hz}$ over a wall-clock interval $T$ of at least 300 s starting at time $t_0$. This allowed us to compute the energy per particle-update,

\begin{equation}
    \epsilon = \frac{E(t_0 + T) - E(t_0)}{\sum_{j=1}^{M_T} \sum_{r=1}^R N_{r,j}},
\end{equation}

\noindent where $N_{r,j}$ is as in Eq.~\eqref{eq:FOM} and $M_T$ is the number of completed timesteps over duration $T$. The value of $E(t)$ was obtained using Intel's Running Average Power Limit (RAPL) PSys counter on Aurora and HPE Cray's \texttt{pm\_counter} on Frontier, which measure energy used by the host CPUs, memory, and all GPUs. Polaris does not provide a comparable energy counter. 

Figure~\ref{fig:node-energy-gain-combined} shows the ratio $\epsilon_{\rm CPU}/\epsilon_{\rm GPU}$ as a function of ppc. GPU execution requires less energy per particle-update over the entire ppc range, with Aurora showing a larger CPU-to-GPU energy ratio than Frontier until 256 ppc in 2D. At 256 ppc, the ratios are approximately $10$ on Aurora and Frontier, corresponding to energy reductions of approximately $90\%$.

\subsection{Application: Parallel collisionless shocks} \label{sec:application}

\begin{figure*}[pos=t]
\centering
\includegraphics[width=\textwidth]{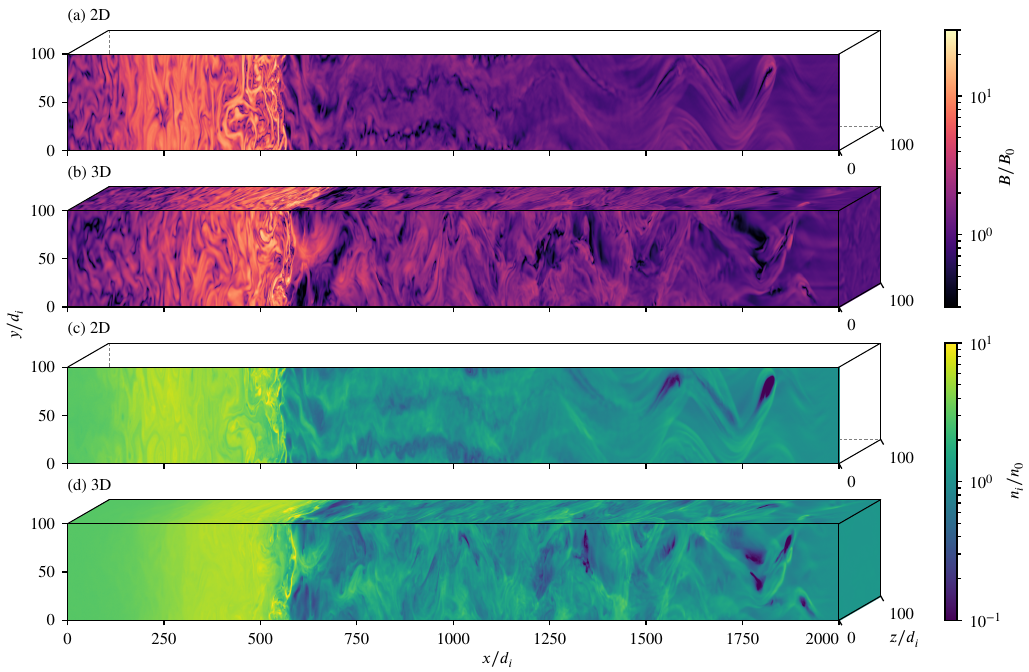}
\caption{\label{fig:collisionless-shock-2d-vs-3d}Magnetic field magnitude $B/B_0$ (a,b) and ion number density $n_i/n_0$ (c,d) at $t\Omega_i \simeq 104$ in 2D and 3D, shown in oblique projection. Ions are continuously injected at $x = 2000\,d_i$ with bulk velocity $\simeq -20\,v_{A0}$ along $x$, reflect off the wall at $x = 0$, and form a collisionless shock located at $x \approx 560\,d_i$ in this snapshot.}
\end{figure*}

To demonstrate the capability of the GPU-accelerated \texttt{dHybridR}, we performed twin 2D and 3D simulations of a benchmark parallel collisionless shock following the setup of Ref.~\cite{Caprioli+2014}. A single proton species uniformly fills the $2000\,d_i\times100\,d_i$ (2D) and $2000\,d_i\times100\,d_i\times100\,d_i$ (3D) domains at density $n_0$, with velocity components sampled from a Maxwellian with drift velocity $\vb{v}_d = -v_d\,\vu{x} \simeq -20\,v_{A0}\,\vu{x}$, antiparallel to the background field $\vb{B}_0 = B_0\vu{x}$, and thermal spread $v_{\rm th} = \sqrt{k_B T_i/m_p} = v_{A0}$. Moreover, this distribution is continuously injected through the $x = 2000\,d_i$ transverse plane. Electrons follow a $\gamma_e = 5/3$ polytropic equation of state with $T_e = T_i$ initially, and $c = 100\,v_{A0}$. The domain is periodic in the transverse directions, open at the injection plane, and reflecting at $x = 0$; ions then immediately begin reflecting off the $x=0$ wall, forming a collisionless shock that propagates along $+x$ into the incoming flow. In the simulation frame, the downstream plasma is at rest and the shock moves at $v_{\rm sh}\,\vu{x}$, whereas in the shock frame, the upstream and downstream plasma move at $\vb{u}_1 = -(v_d + v_{\rm sh})\,\vu{x}$ and $\vb{u}_2 = -v_{\rm sh}\,\vu{x}$, respectively. Mass conservation across a steady planar shock, $n_1 u_1 = n_2 u_2$, then gives the shock-frame Alfv\'enic Mach number \cite{Haggerty+2020}
\begin{equation}\label{eq:shock-frame-mach}
    \mathcal{M}_A = \frac{u_1}{v_{A0}} = \frac{r}{r-1}\,\frac{v_d}{v_{A0}} \approx 25,
\end{equation}
\noindent with $r = n_2/n_1 \approx 4$--$5$ the compression ratio measured in the simulations. The sonic Mach number is comparable to $\mathcal{M}_A$.
Both runs use 5 cells per $d_i$, 64 ppc, and $\Delta t\, \Omega_i = 5.8\times10^{-4}$ across 180,000 steps, resulting in a simulation duration of $t\Omega_i \simeq 104$. A snapshot of both runs at this moment is shown in Fig.~\ref{fig:collisionless-shock-2d-vs-3d}.

Even at this early stage of the shock's evolution, significant differences exist between the runs due to dimensionality. The 2D run does not partition the box-integrated magnetic energy, $U_B = \int B^2/8\pi \,\dd[D]{x}$, equally across the field components: by $t\Omega_i \simeq 100$ the out-of-plane $B_z$ holds about $50\%$ of the magnetic energy and $B_x$, $B_y$ about $25\%$ each, whereas in the 3D run the three components are near equipartition. Likewise, the 3D magnetic field and ion density profiles peak just behind the shock front and taper off downstream, whereas in 2D they remain near their maximum for hundreds of $d_i$ behind it. Since the runs only differ by dimensionality, both are consequences of the artificial translational symmetry in 2D \cite{Jokipii+1993, Giacalone+1994, Jones+1998, Guo+2013, Zacharegkas+2024} and illustrate why 3D runs are necessary to properly study collisionless shocks. 

The 3D simulation took 15.6 h on 512 Frontier nodes ($8.0\times10^{3}$ node-hours), sustaining 86\% of the per-GCD throughput of the 64-ppc full-node Frontier run underlying Fig.~\ref{fig:cpu-gpu-speedup}(c), even though, unlike that benchmark, it includes particle injection, dynamic load balancing, and I/O. On the same nodes' CPUs, the run would require an order of magnitude more node-hours (resulting in a wall-clock time of about a week) [Fig.~\ref{fig:cpu-gpu-speedup}(c)]. Following the shock for the $10^3\,\Omega_i^{-1}$ required for diffusive shock acceleration \cite{Caprioli+2014} requires both longer simulation durations and longer boxes, multiplying this cost further. The original implementation, used in all \texttt{dHybridR} studies prior to Ref.~\cite{Caprioli+2025}, is another $4\times$ slower at 64 ppc on Aurora [Fig.~\ref{fig:cpu-scaling}(b)]. Relative to it, the reduction is $40$--$50\times$ for this run and up to two orders of magnitude at 256 ppc, making such simulations feasible.

\section{Conclusion}\label{sec:conclusion}

We have presented a performance-portable GPU implementation of the hybrid PIC code \texttt{dHybridR} with demonstrated performance across all major GPU vendors. The implementation builds on initial CPU optimizations to the Fortran codebase and employs OpenMP target offload with specialized SYCL kernels for targeted kernel optimization, all in a single codebase.

Thermal-plasma benchmarks in 2D and 3D demonstrate weak scaling efficiencies of $86$--$97\%$ at the largest tested accelerator counts, reaching 49,152 accelerators on each of the exascale systems Aurora and Frontier. At 256 ppc, full-node GPU execution is $14$--$32\times$ faster than the refactored CPU implementation and uses approximately $90\%$ less energy per particle-update on Aurora and Frontier. To our knowledge, no other hybrid PIC code has reported GPU performance at this scale, leaving \texttt{dHybridR} uniquely positioned to exploit such exascale systems. Moreover, our approach requires only OpenMP offload, SYCL, and their interoperability on the target GPU architecture, presenting a practical GPU-porting strategy for existing MPI-parallelized Fortran plasma codes.

Several avenues exist for further optimization. Particle sorting could exploit the largely preserved cell ordering between timesteps, while kernel fusion or graph-based execution could reduce kernel launch overhead. Accumulating fewer moments simultaneously in 3D deposition could additionally lower register pressure and increase occupancy. Such optimizations would support increasingly large and long-duration studies of space and astrophysical plasmas.

\section*{Acknowledgments}

The authors thank Daniel Kokron for his early work on optimizing \texttt{dHybridR} and Colby Haggerty for making that collaboration possible. We also thank the mentors of the Argonne Leadership Computing Facility's 2026 INCITE GPU Hackathon, who suggested SYCL and helped diagnose bottlenecks in the initial OpenMP offload port. D.C. was partially supported by NASA grant 80NSSC18K1726 and NSF grant AST-2510951. This research used resources of the Argonne Leadership Computing Facility, which is a U.S. Department of Energy Office of Science User Facility operated under contract No. DE-AC02-06CH11357, and resources of the Oak Ridge Leadership Computing Facility at the Oak Ridge National Laboratory, which is supported by the Advanced Scientific Computing Research programs in the Office of Science of the U.S. Department of Energy under Contract No. DE-AC05-00OR22725.

\appendix

\section{CPU memory hierarchy, vector instructions, and CPU codebase refactor} \label{app:cpu-refactor}

Modern CPU architectures access data through a \textit{hierarchy of memory levels}, ranging from registers located inside each CPU core, through the on-chip L1, L2, and L3 caches, to main memory. As an example, Table~\ref{tab:memhier} summarizes the properties of the memory hierarchy for the CPUs on an Aurora compute node. Reading data from registers has the lowest latency, which increases down the memory hierarchy through the L1, L2, and L3 caches to main memory. This motivates refactoring code to reuse data already held in registers and caches. 

Although individual memory accesses may address as little as a single byte, the cache hierarchy stores and transfers data in fixed-size blocks called \textit{cache lines}. On x86 processors like the Intel Xeon Max 9470C of Table~\ref{tab:memhier}, a cache line is typically 64 bytes and \textit{aligned}, meaning it begins at an address that is a multiple of 64. When a program requests a value to be loaded from memory, the L1 cache is first checked to verify whether it contains the cache line containing the value. If so, this is deemed an L1 \textit{cache hit}, and if not, an L1 \textit{cache miss}. If a cache miss occurs, storage further down the memory hierarchy is checked until either a cache hit occurs or a 64-byte-aligned region containing the value is ultimately pulled from main memory. 

Alongside the memory hierarchy, modern CPUs provide vector instruction-set extensions that allow one instruction (addition, multiplication, etc.) to operate on multiple values simultaneously. This form of data-level parallelism is known as \textit{single instruction, multiple data} (SIMD). Aurora's Intel Xeon Max 9470C supports AVX-512, whose 512-bit vector registers can each hold 16 single-precision floating-point values. A compiler can therefore auto-vectorize a loop by generating vector instructions that operate on up to 16 single-precision values at once. This is most efficient when loop iterations access adjacent values in memory: 16 consecutive single-precision values span 64 bytes, matching both the width of an AVX-512 register and the size of a cache line, so an aligned block can be read from a single cache line and operated on with a single vector operation. 

Before the refactor, a given species' $S$ properties were stored in one $S\times N$ array, \texttt{p(S,N)}, with $N$ the number of macroparticles of that species. Since Fortran stores arrays in column-major order, the $S$ properties of one macroparticle were then adjacent in memory while the same property of consecutive macroparticles, e.g. \texttt{p(1,i)} and \texttt{p(1,i+1)}, sat $S$ elements apart. A loop over macroparticles that loaded one property, such as position, therefore had a stride of $S$, the \textit{stride} being the distance in elements between the values loaded from memory on consecutive iterations. Only one in $S$ elements of every fetched cache line was used, and the non-unit stride caused the compiler either to keep the loop scalar or, if vectorized, to use costly gather instructions to fill the vector registers rather than simply filling them with a full cache line of adjacent particle data. The electromagnetic field arrays had the same problem, with arrays such as \texttt{B(3,Nx,Ny,Nz)} storing their component ($x$, $y$, or $z$) as the first index. Loops over neighboring grid cells that accessed the same component then ran with a stride of three.

The refactor split a species' property array into $S$ individual one-dimensional arrays: for example, \texttt{x(N)} for the $x$ positions of that species' macroparticles. Similarly, the electromagnetic fields were redefined so that the field component became the last index; for example, \texttt{B(Nx,Ny,Nz,3)}. Loops over macroparticles and grid cells then became unit stride, allowing the compiler to auto-vectorize them with each fetched cache line fully utilized. 

\bibliographystyle{unsrtnat}
\bibliography{references}

@article{Gargate+2007,
    author = {Gargat{\'e}, L. and Bingham, R. and Fonseca, R. A. and Silva, L. O.},
    title = {{dHybrid}: A massively parallel code for hybrid simulations of space plasmas},
    journal = {Computer Physics Communications},
    volume = {176},
    number = {6},
    pages = {419--425},
    year = {2007},
    doi = {10.1016/j.cpc.2006.11.013}
}

@article{Haggerty+2019,
    author = {Haggerty, Colby C. and Caprioli, Damiano},
    title = {{dHybridR}: A Hybrid Particle-in-cell Code Including Relativistic Ion Dynamics},
    journal = {The Astrophysical Journal},
    volume = {887},
    number = {2},
    pages = {165},
    year = {2019},
    doi = {10.3847/1538-4357/ab58c8}
}

@book{Lipatov2002,
    author = {Lipatov, Alexander S.},
    title = {The Hybrid Multiscale Simulation Technology: An Introduction with Application to Astrophysical and Laboratory Plasmas},
    series = {Scientific Computation},
    publisher = {Springer},
    address = {Berlin, Germany},
    year = {2002},
    doi = {10.1007/978-3-662-05012-5},
    isbn = {978-3-540-41734-7}
}

@incollection{Winske+2023,
    author = {Winske, Dan and Karimabadi, Homa and Le, Ari Yitzchak and Omidi, Nojan Nick and Roytershteyn, Vadim and Stanier, Adam John},
    title = {Hybrid-Kinetic Approach: Massless Electrons},
    booktitle = {Space and Astrophysical Plasma Simulation: Methods, Algorithms, and Applications},
    editor = {B{\"u}chner, J{\"o}rg},
    publisher = {Springer},
    address = {Cham, Switzerland},
    year = {2023},
    pages = {63--91},
    doi = {10.1007/978-3-031-11870-8_3},
    isbn = {978-3-031-11869-2}
}

@article{Caprioli+2014,
    doi = {10.1088/0004-637X/783/2/91},
    year = {2014},
    month = feb,
    publisher = {The American Astronomical Society},
    volume = {783},
    number = {2},
    pages = {91},
    author = {Caprioli, D. and Spitkovsky, A.},
    title = {Simulations of ion acceleration at non-relativistic shocks. {I}. {Acceleration} efficiency},
    journal = {The Astrophysical Journal},
}

@article{Jokipii+1993,
    author = {Jokipii, J. R. and K{\'o}ta, J. and Giacalone, J.},
    title = {Perpendicular transport in 1- and 2-dimensional shock simulations},
    journal = {Geophysical Research Letters},
    volume = {20},
    number = {17},
    pages = {1759-1761},
    doi = {10.1029/93GL01973},
    year = {1993}
}

@article{Giacalone+1994,
    author = {Giacalone, J. and Jokipii, J. R.},
    title = {Charged-Particle Motion in Multidimensional Magnetic-Field Turbulence},
    journal = {The Astrophysical Journal},
    volume = {430},
    number = {2},
    pages = {L137--L140},
    year = {1994},
    doi = {10.1086/187457}
}

@article{Jones+1998,
    doi = {10.1086/306480},
    year = {1998},
    month = dec,
    volume = {509},
    number = {1},
    pages = {238},
    author = {Jones, Frank C. and Jokipii, J. Randy and Baring, Matthew G.},
    title = {Charged-Particle Motion in Electromagnetic Fields Having at Least One Ignorable Spatial Coordinate},
    journal = {The Astrophysical Journal},
}

@article{Orusa+2023,
    title = {Fast Particle Acceleration in {3D} Hybrid Simulations of Quasiperpendicular Shocks},
    author = {Orusa, Luca and Caprioli, Damiano},
    journal = {Physical Review Letters},
    volume = {131},
    number = {9},
    pages = {095201},
    numpages = {6},
    year = {2023},
    month = sep,
    publisher = {American Physical Society},
    doi = {10.1103/PhysRevLett.131.095201},
}

@article{Orusa+2026,
    doi = {10.3847/1538-4357/ae563e},
    year = {2026},
    month = apr,
    publisher = {The American Astronomical Society},
    volume = {1001},
    number = {2},
    pages = {158},
    author = {Orusa, Luca and Caprioli, Damiano and Sironi, Lorenzo and Spitkovsky, Anatoly},
    title = {The Role of Three-dimensional Effects on Ion Injection and Acceleration in Perpendicular Shocks},
    journal = {The Astrophysical Journal},
}

@article{Haggerty+2020,
    doi = {10.3847/1538-4357/abbe06},
    year = {2020},
    month = dec,
    publisher = {The American Astronomical Society},
    volume = {905},
    number = {1},
    pages = {1},
    author = {Haggerty, Colby C. and Caprioli, Damiano},
    title = {Kinetic Simulations of Cosmic-Ray-modified Shocks. {I}. {Hydrodynamics}},
    journal = {The Astrophysical Journal},
}

@article{Le+2023,
    author = {Le, Ari and Stanier, Adam and Yin, Lin and Wetherton, Blake and Keenan, Brett and Albright, Brian},
    title = {{Hybrid-VPIC}: An open-source kinetic/fluid hybrid particle-in-cell code},
    journal = {Physics of Plasmas},
    volume = {30},
    number = {6},
    pages = {063902},
    year = {2023},
    month = jun,
    issn = {1070-664X},
    doi = {10.1063/5.0146529},
}

@article{Zacharegkas+2024,
    doi = {10.3847/1538-4357/ad3960},
    year = {2024},
    month = may,
    publisher = {The American Astronomical Society},
    volume = {967},
    number = {1},
    pages = {71},
    author = {Zacharegkas, Georgios and Caprioli, Damiano and Haggerty, Colby and Gupta, Siddhartha and Schroer, Benedikt},
    title = {Modeling the Saturation of the {Bell} Instability Using Hybrid Simulations},
    journal = {The Astrophysical Journal},
}

@article{Myers+2021,
    title = {Porting {WarpX} to {GPU}-accelerated platforms},
    journal = {Parallel Computing},
    volume = {108},
    pages = {102833},
    year = {2021},
    issn = {0167-8191},
    doi = {10.1016/j.parco.2021.102833},
    author = {A. Myers and A. Almgren and L.D. Amorim and J. Bell and L. Fedeli and L. Ge and K. Gott and D.P. Grote and M. Hogan and A. Huebl and R. Jambunathan and R. Lehe and C. Ng and M. Rowan and O. Shapoval and M. Thévenet and J.-L. Vay and H. Vincenti and E. Yang and N. Zaïm and W. Zhang and Y. Zhao and E. Zoni},
}

@article{Lee+2025, 
    title={Acceleration of the particle-in-cell code {Osiris} with graphics processing units}, volume={91}, 
    DOI={10.1017/S0022377824001569}, 
    number={1}, 
    journal={Journal of Plasma Physics}, author={Lee, Roman P. and Pierce, Jacob R. and Miller, Kyle G. and Almanza, Maria and Tableman, Adam and Decyk, Viktor K. and Fonseca, Ricardo A. and Alves, E. Paulo and Mori, Warren B.}, 
    year={2025}, 
    pages={E8},
}

@ARTICLE{Bird+2022,
    author = {Bird, Robert and Tan, Nigel and Luedtke, Scott V. and Harrell, Stephen Lien and Taufer, Michela and Albright, Brian},
    journal = {IEEE Transactions on Parallel and Distributed Systems}, 
    title = {{VPIC} 2.0: Next Generation Particle-in-Cell Simulations}, 
    year = {2022},
    volume = {33},
    number = {4},
    pages = {952-963},
    doi = {10.1109/TPDS.2021.3084795},
}

@article{Hakobyan+2026,
    doi = {10.3847/1538-4365/ae6591},
    year = {2026},
    month = jun,
    publisher = {The American Astronomical Society},
    volume = {285},
    number = {1},
    pages = {11},
    author = {Hakobyan, Hayk and Böss, Ludwig M. and Cai, Yangyang and Chernoglazov, Alexander and Galishnikova, Alisa and Gorbunov, Evgeny A. and Mahlmann, Jens F. and Philippov, Alexander and Solanki, Siddhant and Vanthieghem, Arno and Zhou, Muni and {(Entity Development Team)}},
    title = {Entity—Hardware-agnostic Particle-in-cell Code for Plasma Astrophysics. {I}. {Curvilinear} Special Relativistic Module},
    journal = {The Astrophysical Journal Supplement Series},
}

@article{Fatemi+2017,
    doi = {10.1088/1742-6596/837/1/012017},
    year = {2017},
    month = may,
    publisher = {IOP Publishing},
    volume = {837},
    number = {1},
    pages = {012017},
    author = {Fatemi, Shahab and Poppe, Andrew R. and Delory, Gregory T. and Farrell, William M.},
    title = {{AMITIS}: A {3D} {GPU}-Based Hybrid-{PIC} Model for Space and Plasma Physics},
    journal = {Journal of Physics: Conference Series},
}

@Article{Behar+2022,
    AUTHOR = {Behar, E. and Fatemi, S. and Henri, P. and Holmstr\"om, M.},
    TITLE = {{Menura}: a code for simulating the interaction between a turbulent solar wind and solar system bodies},
    JOURNAL = {Annales Geophysicae},
    VOLUME = {40},
    YEAR = {2022},
    NUMBER = {3},
    PAGES = {281--297},
    DOI = {10.5194/angeo-40-281-2022},
}

@article{Kim+2025,
    title = {Hybrid {EPIC-GOD}: An energy-conserving hybrid particle-in-cell code for {GPU} acceleration using {OpenACC}},
    journal = {Computer Physics Communications},
    volume = {315},
    pages = {109726},
    year = {2025},
    issn = {0010-4655},
    doi = {10.1016/j.cpc.2025.109726},
    author = {Sunjung Kim and Dongsu Ryu and G.S. Choe and Sibaek Yi},
}

@article{Groenewald+2023,
    author = {Groenewald, R. E. and Veksler, A. and Ceccherini, F. and Necas, A. and Nicks, B. S. and Barnes, D. C. and Tajima, T. and Dettrick, S. A.},
    title = {Accelerated kinetic model for global macro stability studies of high-beta fusion reactors},
    journal = {Physics of Plasmas},
    volume = {30},
    number = {12},
    pages = {122508},
    year = {2023},
    month = dec,
    issn = {1070-664X},
    doi = {10.1063/5.0178288},
}

@article{Harned1982,
    title = {Quasineutral hybrid simulation of macroscopic plasma phenomena},
    journal = {Journal of Computational Physics},
    volume = {47},
    number = {3},
    pages = {452-462},
    year = {1982},
    issn = {0021-9991},
    doi = {10.1016/0021-9991(82)90094-8},
    author = {Douglas S. Harned},
}

@article{Kunz+2014,
    title = {Pegasus: A new hybrid-kinetic particle-in-cell code for astrophysical plasma dynamics},
    journal = {Journal of Computational Physics},
    volume = {259},
    pages = {154-174},
    year = {2014},
    issn = {0021-9991},
    doi = {10.1016/j.jcp.2013.11.035},
    author = {Matthew W. Kunz and James M. Stone and Xue-Ning Bai},
}

@article{Stanier+2019,
    title = {A fully implicit, conservative, non-linear, electromagnetic hybrid particle-ion/fluid-electron algorithm},
    journal = {Journal of Computational Physics},
    volume = {376},
    pages = {597-616},
    year = {2019},
    issn = {0021-9991},
    doi = {10.1016/j.jcp.2018.09.038},
    author = {A. Stanier and L. Chacón and G. Chen},
}

@article{Ripperda+2018,
    doi = {10.3847/1538-4365/aab114},
    year = {2018},
    month = mar,
    publisher = {The American Astronomical Society},
    volume = {235},
    number = {1},
    pages = {21},
    author = {Ripperda, B. and Bacchini, F. and Teunissen, J. and Xia, C. and Porth, O. and Sironi, L. and Lapenta, G. and Keppens, R.},
    title = {A Comprehensive Comparison of Relativistic Particle Integrators},
    journal = {The Astrophysical Journal Supplement Series},
}

@book{Hockney1988,
    author = {Hockney, Roger W. and Eastwood, James W.},
    title = {Computer Simulation Using Particles},
    publisher = {CRC Press},
    year = {1988},
    doi = {10.1201/9780367806934}
}

@inproceedings{Boris1970,
    author = {Boris, J. P.},
    title = {Relativistic Plasma Simulation---Optimization of a Hybrid Code},
    booktitle = {Proceedings of the Fourth Conference on Numerical Simulation of Plasmas},
    editor = {Boris, J. P. and Shanny, R. A.},
    publisher = {Naval Research Laboratory},
    address = {Washington, D.C.},
    pages = {3--67},
    year = {1970}
}

@article{Bowers2001,
    title = {Accelerating a Particle-in-Cell Simulation Using a Hybrid Counting Sort},
    journal = {Journal of Computational Physics},
    volume = {173},
    number = {2},
    pages = {393-411},
    year = {2001},
    issn = {0021-9991},
    doi = {10.1006/jcph.2001.6851},
    author = {K. J. Bowers},
}

@article{Williams+2009,
    author = {Williams, Samuel and Waterman, Andrew and Patterson, David},
    title = {Roofline: an insightful visual performance model for multicore architectures},
    year = {2009},
    issue_date = {April 2009},
    publisher = {Association for Computing Machinery},
    address = {New York, NY, USA},
    volume = {52},
    number = {4},
    issn = {0001-0782},
    doi = {10.1145/1498765.1498785},
    journal = {Communications of the ACM},
    month = apr,
    pages = {65--76},
    numpages = {12}
}

@article{Beck+2019,
    title = {Adaptive {SIMD} optimizations in particle-in-cell codes with fine-grain particle sorting},
    journal = {Computer Physics Communications},
    volume = {244},
    pages = {246-263},
    year = {2019},
    issn = {0010-4655},
    doi = {10.1016/j.cpc.2019.05.001},
    author = {A. Beck and J. Derouillat and M. Lobet and A. Farjallah and F. Massimo and I. Zemzemi and F. Perez and T. Vinci and M. Grech},
}

@book{Birdsall+1991,
  author = {Birdsall, Charles K. and Langdon, A. Bruce},
  title = {Plasma Physics via Computer Simulation},
  publisher = {CRC Press},
  year = {1991},
  doi = {10.1201/9781315275048},
  isbn = {9781315275048}
}

@article{Melzani+2013,
    author = {Melzani, Micka{\"e}l and Winisdoerffer, Christophe and Walder, Rolf and Folini, Doris and Favre, Jean M. and Krastanov, Stefan and Messmer, Peter},
    title = {{Apar-T}: code, validation, and physical interpretation of particle-in-cell results},
    doi= "10.1051/0004-6361/201321557",
    journal = {Astronomy \& Astrophysics},
    year = 2013,
    volume = 558,
    pages = "A133",
}

@article{Zenitani2015,
    author = {Zenitani, Seiji},
    title = {Loading relativistic {Maxwell} distributions in particle simulations},
    journal = {Physics of Plasmas},
    volume = {22},
    number = {4},
    pages = {042116},
    year = {2015},
    month = apr,
    issn = {1070-664X},
    doi = {10.1063/1.4919383},
}

@ARTICLE{Yee1966,
    author = {Kane Yee},
    journal = {IEEE Transactions on Antennas and Propagation}, 
    title = {Numerical solution of initial boundary value problems involving {Maxwell's} equations in isotropic media}, 
    year = {1966},
    volume = {14},
    number = {3},
    pages = {302-307},
    doi = {10.1109/TAP.1966.1138693}
}

@article{Nielson+1976,
    title = {Particle-code models in the nonradiative limit},
    author = {Nielson, Clair W. and Lewis, H. Ralph},
    journal = {Methods in Computational Physics},
    volume = {16},
    pages = {367--388},
    year = {1976}
}

@article{Hewett1985,
    author = {Hewett, D. W.},
    title = {Elimination of electromagnetic radiation in plasma simulation: the {D}arwin or magnetoinductive approximation},
    journal = {Space Science Reviews},
    volume = {42},
    pages = {29--40},
    year = {1985},
    doi = {10.1007/BF00218221}
}

@inproceedings{Fedeli+2022,
    author = {Fedeli, Luca and Huebl, Axel and Boillod-Cerneux, France and Clark, Thomas and Gott, Kevin and Hillairet, Conrad and Jaure, Stephan and Leblanc, Adrien and Lehe, R{\'e}mi and Myers, Andrew and Piechurski, Christelle and Sato, Mitsuhisa and Zaim, Ne{\"i}l and Zhang, Weiqun and Vay, Jean-Luc and Vincenti, Henri},
    booktitle = {SC22: International Conference for High Performance Computing, Networking, Storage and Analysis}, 
    title = {Pushing the Frontier in the Design of Laser-Based Electron Accelerators with Groundbreaking Mesh-Refined Particle-In-Cell Simulations on Exascale-Class Supercomputers}, 
    year = {2022},
    pages = {1-12},
    doi = {10.1109/SC41404.2022.00008}
}

@article{Guo+2013,
    doi = {10.1088/0004-637X/773/2/158},
    year = {2013},
    month = aug,
    publisher = {The American Astronomical Society},
    volume = {773},
    number = {2},
    pages = {158},
    author = {Guo, Fan and Giacalone, Joe},
    title = {THE ACCELERATION OF THERMAL PROTONS AT PARALLEL COLLISIONLESS SHOCKS: THREE-DIMENSIONAL HYBRID SIMULATIONS},
    journal = {The Astrophysical Journal},
}

@misc{Intel-GPU-optimization-guide,
    author = {{Intel Corporation}},
    title = {{oneAPI GPU Optimization Guide}: {Intel Xe GPU Architecture}},
    year = {2025},
    howpublished = {\url{https://www.intel.com/content/www/us/en/docs/oneapi/optimization-guide-gpu/2025-2/intel-xe-gpu-architecture.html}},
    note = {Version 2025.2, Doc.\ ID 771772, 10 July 2025; archived at \url{https://archive.ph/cv7OM}}
}

@misc{Intel-GPU-occupancy-guide,
    author = {{Intel Corporation}},
    title = {{oneAPI GPU Optimization Guide}: {Thread Mapping and GPU Occupancy}},
    year = {2025},
    howpublished = {\url{https://www.intel.com/content/www/us/en/docs/oneapi/optimization-guide-gpu/2025-2/thread-mapping-and-gpu-occupancy.html}},
    note = {Version 2025.2, Doc.\ ID 771772, 10 July 2025; archived at \url{https://archive.ph/hqtdk}}
}

@article{Valentini+2007,
    author = {Valentini, F. and Tr{\'a}vn{\'\i}{\v{c}}ek, P. and Califano, F. and Hellinger, P. and Mangeney, A.},
    title = {A hybrid-{Vlasov} model based on the current advance method for the simulation of collisionless magnetized plasma},
    journal = {Journal of Computational Physics},
    year = {2007},
    volume = {225},
    number = {1},
    pages = {753--770},
    doi = {10.1016/j.jcp.2007.01.001}
}

@article{Pezzi+2019,
    author = {Pezzi, O. and Cozzani, G. and Califano, F. and Valentini, F. and Guarrasi, M. and Camporeale, E. and Brunetti, G. and Retin{\`o}, A. and Veltri, P.},
    title = {{ViDA}: a {Vlasov}--{DArwin} solver for plasma physics at electron scales},
    journal = {Journal of Plasma Physics},
    year = {2019},
    volume = {85},
    number = {5},
    pages = {905850506},
    doi = {10.1017/S0022377819000631},
}

@article{Juno+2018,
    author = {Juno, J. and Hakim, A. and TenBarge, J. and Shi, E. and Dorland, W.},
    title = {Discontinuous {Galerkin} algorithms for fully kinetic plasmas},
    journal = {Journal of Computational Physics},
    year = {2018},
    volume = {353},
    pages = {110--147},
    doi = {10.1016/j.jcp.2017.10.009},
}

@article{Palmroth+2018,
    author = {Palmroth, M. and Ganse, U. and Pfau-Kempf, Y. and Battarbee, M. and Turc, L. and Brito, T. and Grandin, M. and Hoilijoki, S. and Sandroos, A. and {von Alfthan}, S.},
    title = {{Vlasov} methods in space physics and astrophysics},
    journal = {Living Reviews in Computational Astrophysics},
    year = {2018},
    volume = {4},
    pages = {1},
    doi = {10.1007/s41115-018-0003-2}
}

@article{Ganse+2023,
    author = {Ganse, Urs and Koskela, Tuomas and Battarbee, Markus and Pfau-Kempf, Yann and Papadakis, Konstantinos and Alho, Markku and Bussov, Maarja and Cozzani, Giulia and Dubart, Maxime and George, Harriet and Gordeev, Evgeny and Grandin, Maxime and Horaites, Konstantinos and Suni, Jonas and Tarvus, Vertti and Kebede, Fasil Tesema and Turc, Lucile and Zhou, Hongyang and Palmroth, Minna},
    title = {Enabling technology for global {3D} + {3V} hybrid-{Vlasov} simulations of near-{Earth} space},
    journal = {Physics of Plasmas},
    volume = {30},
    number = {4},
    pages = {042902},
    year = {2023},
    month = apr,
    issn = {1070-664X},
    doi = {10.1063/5.0134387},
}

@article{Caprioli+2025,
    doi = {10.3847/2041-8213/ae109b},
    year = {2025},
    month = oct,
    publisher = {The American Astronomical Society},
    volume = {993},
    number = {1},
    pages = {L1},
    author = {Caprioli, Damiano and Orusa, Luca and Cernetic, Miha and Haggerty, Colby C. and Ostler, Bricker},
    title = {Acceleration of Heavy Ions at Nonrelativistic Collisionless Shocks},
    journal = {The Astrophysical Journal Letters},
}

\end{document}